\documentclass{article}       

\usepackage{graphicx}
\begin{document}
	
\title{On the noise in statistics of PIV measurements}

\author{William K. George$^1$   \and
	Michel Stanislas$^2$}

\maketitle

\begin{center}
\noindent$^1$ Visiting Pr., Centrale Lille, F59651 Villeneuve d'Ascq, France\\
$^2$ Pr. Emeritus, Centrale Lille, F59651 Villeneuve d'Ascq, France\\
corresponding author: michel.stanislas@ec-lille.fr
\end{center}
\medskip

\begin{abstract}
It is argued herein that when PIV is used to measure turbulence, it can be treated as a time-dependent signal.  The `output' velocity consists of three primary contributions:  the time-dependent velocity, a noise arising from the quantization (or pixelization), and a noise contribution from the fact that the velocity is not uniform inside the interrogation volume.  For both of the latter their variances depend inversely on the average number of particles or images in this interrogation volume.  All three of these are spatially filtered by the finite extent of the interrogation window.  Since the above noises are associated directly with the individual particles (or particle images), the noise between different realizations and different interrogation volumes is statistically independent.
\end{abstract}

\section*{List of Symbols}

\noindent
\begin{tabular}{ll}
	$\vec{a}$ & Lagrangian particle coordinate \\
	$b_{cam}(\vec{x}_s,t)$ & Particle image on detector \\
	$B_{oi,j}(\vec{y},\vec{y'})$ &Two-points single-time velocity correlation tensor\\
	$c(\vec{y},t)$ & Image of all particles inside the interrogation volume \\
	$C_o$        & Correlation of images\\
	$E_{11}(k_1)$&One-dimensional velocity fluctuation spectrum\\
	$E_{noise}(k_1)$&Noise one-dimensional spectrum\\
	$f(\vec{x},t)$&Particle image back-projected in physical space\\
	$f$&Lens focal length\\
	$f$\#&Lens f-number\\
	$F_{oi,j}(\vec{k})$&Measured three-dimensional spectrum\\
	$\tilde{F}_{i,j}(\vec{k})$&Volume-averaged three-dimensional spectrum\\
	$F_{oi,j}^1(\vec{k_1})$&Measured one-dimensional spectrum\\
	$\tilde{F}_{i,j}^1(\vec{k_1})$&Volume-averaged one-dimensional spectrum\\
	$F_{i,j}^1(\vec{k_1})$&True one-dimensional spectrum\\
	$F_{noise}^1(\vec{k_1})$&Noise one-dimensional spectrum\\
	$g(\vec{a})$&Particle location function: $\delta(\vec{a})$\\
	$I = 1.492$&Integral over $[0,2\pi]$ of the sinc function\\
	$\vec{k}$&Wave vector\\
	$k_i$&Wave number along $x_i$\\
	$M$& Magnification of recording system\\
	$n(\vec{y},t)$&Number of particles inside the interrogation volume\\
	$N(\vec{y})$&Average number of particles inside the interrogation volume\\
	$N_p$& Number of particle per pixel (ppp)\\
	$P_i$&Uncertainty on particle position due to pixelisation\\
	$\vec{r}$&Position vector inside the interrogation volume\\
	$t$&Time\\
	$t_o$&Initial time\\
	$U$&Mean velocity along $x$\\
	$u_i(\vec{x},t)$&Instantaneous local velocity component\\
	$u'_i(\vec{x},t)$&Instantaneous local velocity fluctuation component\\
	$\tilde{u}_i(\vec{x},t)$&Instantaneous volume-averaged velocity component\\
	$u_{oi}(\vec{x},t)$&Correlation-estimated instantaneous velocity component\\
	$\Delta u_{oi}(\vec{x},t) = u_{oi}(\vec{x},t) - \tilde{u}_i(\vec{x},t) $&Instantaneous velocity component error\\
	$<u'_iu'_j>$&Single-point Reynolds stress tensor\\
	$<u'^2_1>^{vol}$&Volume-averaged turbulence intensity\\
	$v_i(\vec{a},t)$&Individual particle Lagrangian velocity\\
	$V(\vec{y})$&Volume of the interrogation volume\\
	$w(\vec{y},t)$&Interrogation volume\\
	$W(\vec{y},\vec{y'})$&Triangular window function\\
	$W_1(U\delta t)$&Window overlap parameter\\
	$\hat{W}(\vec{k})$&Fourier transform of $W(\vec{y},\vec{y'})$\\
	$\vec{x}$&Location vector on physical space\\
	$x_i$&Coordinates in physical space, also $(x,y,z)$\\
	$\vec{X}(\vec{a},t)$&Lagrangian displacement field\\
	$\vec{x}_s$&Position vector on the detector\\
	$X,Y$&Dimensions of interrogation window (also $\Delta_1,\Delta_2$)\\
	$\vec{y}$&Position vector of the center of the interrogation volume\\
	
\end{tabular} 

\noindent
\begin{tabular}{ll}
	$\Delta$&Pixel size\\
	$\Delta_i$&Dimension of the interrogation volume along $x_i$\\
	$\delta_i(\vec{y},\vec{y'}) = u_{oi}(\vec{y},t) - u_{oi}(\vec{y'},t)$& Measured velocity difference between $\vec{y}$ and $\vec{y'}$\\
	$\delta_i^{noise}(\vec{y},\vec{y'}) = \Delta u_{oi}(\vec{y},t) - \Delta u_{oi}(\vec{y'},t)$& Noise difference between $\vec{y}$ and $\vec{y'}$\\
	$\epsilon_{ik} = <\delta_i \delta_k>$&Second moment of velocity differences\\
	$\zeta$&Noise characteristic constant\\	
	$<\eta^2>$&Camera noise term\\
	$\mu$&Number of particles per unit volume\\
	$\xi_{ik}$&Noise on $\epsilon_{ik}$\\
	$\sigma$&Noise variance\\
\end{tabular} 

\section{Introduction}

A demonstrative application of Particle Image Velocimetry (PIV) was first proposed to the scientific community as "Speckle Velocimetry" by \cite{meynart83} in 1983. As with all measurement techniques, its theoretical assesment, with the aim of quantifying its accuracy, was developed progressively in parallel to its technical progress. What is specific to PIV, compared to other measurement techniques in fluid mechanics, is the fact that the technology has undergone several revolutions that have affected strongly its characteristics (and this is most probably not finished). Starting from a double-pulsed ruby laser with one set of pulses every 30s, photographic film recording and optical Fourier transform in 1983, we are now using reliable high repetition rate Nd-YAG lasers with low noise high sensitivity CCD or CMOS cameras having millions of pixels, and all the image processing down to the output of velocity vector fields is done in nearly real time even for 3D3C (3 dimensions-3 components) measurements. This can be compared to Hot Wire Anemometry or Laser Doppler Anemometry, for example, for which the configuration was stabilized after about 20 years and the development abandonned progressively by researchers. PIV has been since the beginning  (and still is 40 years later) a field of continuous successive technical improvements and impressive creativity of the research community. It is the combination of the availability of always new and better devices, of the progress of the theory and of the inventivity of the scientists which explains mostly this tremendous evolution. 

The aim in the present contribution is not to make a historical review of the development of the technique over the years, nor of its theory. The early development of the method is quite well covered by \cite{grant94} and the present status of the theory is well developed in the book of R.J. Adrian and J. Westerweel who are two  pioneers of its development \cite{adrianwesterweel2011} and the one of the DLR team \cite{raffel2011} which was also  the originator of many significant improvements of the technique. 

Beside significant theoretical developments, much understanding has been also gained thanks to the construction of sophisticated synthetic image generators. A good example of such a tool is the generator developed in the frame of the EUROPIV European project \cite{lecordier04}. Its use  is well illustrated for example by the contribution of \cite{foucaut04a}, but also by the extensive work performed by the scientific community in the frame of the successive PIV challenges organized by the PIVNET European network \cite{stanislas03,stanislas05a,stanislas08}.  Finally, several authors have looked at an ``a posteriori'' estimation of the measurement error, see for example \cite{wieneke15}. Thanks to all these contributions, the PIV technique is quite well understood from the point of view of its theory, and it is possible on this basis to optimize a set-up in order to get the best out of the measurements. Nevertheless, as it is a complex technique involving many different components and physical principles, some routes for further progress are still open such as the scattering characteristics of the particles, the effect of the coherence of the laser light on the particle images shape and the transfer function of the camera, to cite only some of them. 

At the early stage of its development PIV could not be considered as an adequate tool to assess turbulence. Limited spatial resolution, limited dynamic range, limited number of samples made it impossible to compete with the Hot Wire Anemometer (HWA) which was the reference tool in turbulence, especially when looking at spectra. But the tremendous improvements encountered by the technique over the course of years have allowed progressively researchers to show measurements of high quality of turbulence statistics with the additional benefit of having access at the same time to the instantaneous turbulence structure \cite{adrian00,carlier05}. Even though the spatial resolution and the number of samples have been greatly improved, the dynamic range of PIV compared to HWA is still a concern; and accurate turbulence statistics measurements require a careful optimization of the recording parameters \cite{foucaut04}. 

The aim of the present work is to acquire a better understanding of the noise contribution to the statistics of turbulence measured by PIV. For that purpose, focus is not on the technical limitations or imperfections of the devices such as laser beam profile or pixel size, etc. Attention is back to the basics. First the fact that what PIV measures is the instantaneous `averaged' velocity of a set (usually about 10) of particles randomly distributed inside a small interrogation volume. And second the fact that the image is digitized before any processing is applied to it. So in this paper, the PIV tool will be considered perfect: the particles follow exactly the flow, the particle images are all the same and perfectly Gaussian, and the camera is perfectly pixelizing the image. What will be investigated is how far the result provided by this perfect tool is from the instantaneously averaged velocity of the fluid inside an interrogation volume where the local velocity is not uniform. To do so, we examine  PIV theoretically using the methodologies developed earlier by \cite{GeorgeLumley1973}, \cite{george88}, \cite{Buchhaveetal1979}, and most recently \cite{Velteetal2014} to study Laser Doppler Anemometry. 

As the authors have no possibility to perform an experiment themselves, an extensive use is made of the results of a study of the PIV noise by \cite{foucaut04}. The main interest of this experiment is that measurements were performed both in a fluid at rest and in a turbulent flow with different recording parameters. The analysis of these data allowed the authors to derive an empirical model for the noise and the PIV spectrum. As will be seen, the present theory not only validates this model, but provides also explicit theoretical expressions for the empirical parameters.


\section{Determining the PIV ``output'' \label{sec-image}}

\subsection{A simplified PIV model}
{This section follows almost exactly the formulation used in many papers by Adrian and Westerweel \cite{westerweel13}, and most notably their book~\cite{adrianwesterweel2011}. The primary difference is the use of generalized functions instead of finite (but time-dependent) sums.  It is slightly more general, at least at the beginning, since it uses Lagrangian coordinates in which the particle initial locations can be assumed to be statistically independent of their motion, an assumption generally not valid once turbulence moves them around.}

In our model we  imagine that each particle creates an image which moves across a detector as the particle moves. We
imagine the system to have been perfectly aligned and calibrated,
and all the particles are assumed to create scaled versions of
exactly the same image in the detector-plane, say
$b_{cam}(\vec{x_s},t)$, where $\vec{x_s} = (x_s,y_s)$ is the position vector on the detector. We could as
well include the varying amplitude because of their size and
position within the light beam, but for now we will not.  

By appropriate coordinate transformation, we can link each of these
images to the particle coordinate in the flow, say $\vec{x} = (x,y,z)$.  The
origin of both coordinate systems can be chosen arbitrarily, so
let's choose the origins of $x$ and $x_s$ to coincide, and $y$ and
$y_s$ to coincide.  We choose the origin of $z$ to be in the center of the
illuminating laser sheet.  Since there is assumed to be a one-to-one
mapping between the position of the particle within the beam and its
position on the camera backplane, we can choose instead to work in
the flow coordinate system; i.e., we can write the image produced by
each particle as $f(\vec{x},t)$. 

As each particle moves through the interrogation volume, we assume it follows the
flow so its position is given exactly by the Lagrangian displacement
field, $x_i = X_i[\vec{a},t]$, where $\vec{a}$ is the initial (or
Lagrangian) coordinate at time $t=t_o$ where $t_o$ can be chosen
arbitrarily. We can use this to animate the particle images by
writing them as $f(\vec{X}[\vec{a},t])$ where only those initial
locations which had a particle there can create an image. 

Now to complete this model we need a way to establish which fluid
particles correspond to scattering particles at the initial
instant.
We can `{\it locate}' the fluid particles which have been replaced
by real particles by defining a random locator function, say $g(\vec{a})$,
which turns on where there is a particle and is zero everywhere
else.  We could make $g(\vec{a})$ to have finite dimensions, but it
is easier just to make $g(\vec{a}) = \delta(\vec{a})$ wherever there
is a real particle.  So in this case, $g(\vec{a})$ is really an infinite sum of delta-functions popping up randomly wherever there was a particle in the initial instant.  Note that these could as well have random amplitudes, and thus account for different particle image sizes.

An image of the flow at any instant is the superposition of the
images of all the particles present in the field of view at that
instant.  What happens in practice is that a small portion of the
overall picture is examined at subsequent instants, and the `{\it
instantaneous average velocity'\footnote{`Displacement' and `velocity' are assumed equivalent for the short time intervals of interest herein.}} of the particles present in the
interrogation volume is determined.\footnote{Note that there are several ways the word `average' is used in this discussion.  In this context it means `sum over the particles present at an instant'.  This is of course just a finite sum estimator.  When ensemble average over many independent realizations is meant, we will used the symbols $\langle ~\rangle$, or define a symbol using them.} This is then outputted as the
value most representative of the velocity of that particular
interrogation area at that instant in time.  

The actual procedure by
which this process occurs is in fact not an average at all, but
really a finite sum correlation based on the number of particles
present for both light pulses.  If $w(\vec{y},\vec{r})$ defines our interrogation volume, with $\vec{y}$ the position of the center of the volume and $\vec{r}$ the position vector inside this volume with respect to its center, then the
image due to all the particles inside the interrogation
volume centered at location $\vec{y}$ at time $t$ is given by:

\begin{equation}
c(\vec{y},t) = \int_{as} f(\vec{X}[\vec{a},t]) w(\vec{y},\vec{X}[\vec{a},t]-\vec{y}) g(\vec{a}) d\vec{a}
\label{eq:image}
\end{equation}
where the integral is over all space (abbreviated `{$as$').
$\vec{X}[\vec{a},t]$ is the Lagrangian displacement field of the particle
which started at location $\vec{a}$ at $t=t_o$. Note that the present location of the particle is given by $\vec{x} = \vec{X}[\vec{a},t]$ and the initial position corresponds to 
 $\vec{a} = \vec{X}[\vec{a},t_o]$.
 The $g(\vec{a})$ are generalized functions of the Lagrangian (or initial) coordinate, $\vec{a}$, and turn on only where there is a particle.  Note that in Lagrangian coordinates if there is a particle at $\vec{a}$  at time $t=t_o$, then there is always a particle there.

Using this, we can represent the correlation between two images at $t$ and $t'$ by:
\begin{eqnarray}
& & C_o(\vec{y},\vec{y}',t,t')    =   \langle c(\vec{y},t)~c(\vec{y}',t')\rangle ~ =\nonumber\\
& &  \int \int_{as}
 \langle f(\vec{X}) f(\vec{X'})~ w(\vec{y},\vec{X}-\vec{y})w(\vec{y}',\vec{X'}-\vec{y}')~
 g(\vec{a})g(\vec{a}') d\vec{a}d\vec{a}' \rangle  
\label{eq:imagecorrelation1}
\end{eqnarray}
with $\vec{X} = \vec{X}[\vec{a},t]$ and $\vec{X'} = \vec{X}[\vec{a'},t']$

We will assume at the outset that all flow processes are
stationary random processes, implying that the single time
statistics are time-independent, and that the two-time statistics
depend only on time differences.  This also means that in practice
ensemble averages can be replaced by time averages; or equivalently,
by taking the arithmetic average of data produced over many frames. These equivalent averages will be noted $\langle \rangle$ hereafter

The mean displacement, say $D_i$, of the set of particles in the interrogation volume divided by the time between flashes, say $\delta t$, is identified as the velocity associated with the interrogation volume at the instant between the flashes, $\tilde{t} = t +\delta t/2$.  For future reference we shall refer to this as $u_{oi}(\vec{y},\tilde{t})$ and define $u_{oi}(\vec{y},\tilde{t}) = D_i(\vec{y},\tilde{t})/\delta t$.  Note that we have assumed that $\delta t$ is the same for all realizations. Hereafter we shall drop the tilde and simply refer to this as $u_{oi}(\vec{y},t)$.

The statistics of $g(\vec{a})$ have been previously established  {\it if and only if} all the initial positions are truly random (v. \cite{GeorgeLumley1973,Buchhaveetal1979,Velteetal2014}).  If so,
\begin{eqnarray}
\langle g(\vec{a})g(\vec{a'}) \rangle = \mu^2 +
\mu \delta(\vec{a'}
- \vec{a}) \label{eq:ggprime1}
\end{eqnarray}
  $\langle{g}\rangle = \mu$ represents the number of particles per unit volume.  Or equivalently,
  
  \begin{eqnarray}
  \langle[ g(\vec{a}) - \langle g \rangle] [ g(\vec{a}') - \langle g \rangle] \rangle =
  \mu \delta(\vec{a'}
  - \vec{a}) \label{eq:ggprime1}
  \end{eqnarray}
  This basically says that the locations of any two particles initially are uncorrelated, a direct consequence of their assumed statistically independent locations.    Note that this does not imply that they remain uncorrelated as they are moved by the flow.  This is, of course, the primary reason for choosing to work in Lagrangian coordinates, since only in these can the particles locations be assumed to be statistically independent of their displacements (or velocities).  We note for future reference that the $\mu^2$-contribution is a consequence of multiple particles in the same location at different times.  And it is this contribution that will be discarded below when keeping only the self-product terms in the image correlations.

Given our hypothesis of statistical independence of particle initial location and the fluid motion it follows immediately that the two-time two-space image correlation reduces to:

\begin{eqnarray}
& & C_o(\vec{y},\vec{y}',t,t')     \label{eq:imagecorrelation2} \nonumber\\
 & &  = \mu \int_{as}
 \langle f(\vec{X}) f(\vec{X}) \rangle  w(\vec{y},\vec{X}-\vec{y}) w(\vec{y}',\vec{X}-\vec{y}')d\vec{a} ~ + \nonumber \\
& & \mu^2 \int \int_{as}
\langle f(\vec{X}) f(\vec{X'}) w(\vec{y},\vec{X}-\vec{y})w(\vec{y}',\vec{X'}-\vec{y}')\rangle  d\vec{a}d\vec{a'} 
\end{eqnarray}
The first integral is the self-product which is of primary interest, and the position of this peak  is generally reported as the instantaneous velocity  $u_{oi}(\vec{y},t)$ for a given interrogation volume, as defined above. The second integral contains the base line and much smaller correlations between multiple particles which accidentally coincide, and is generally discarded.  The books by \cite{adrianwesterweel2011} and \cite{raffel2011} discuss in great detail how the peak of the first integral can be analyzed to produce reliable instantaneous velocities and relatively noise-free results.  The most common approach is to approximate the Airy functions of the images by a Gaussian, then fit a 3-point Gaussian to the correlation near the peak.

What is important to us in the remainder of this paper are the following results from these abundant previous efforts:
	
	\begin{enumerate}
		\item The expected values of these image correlations converge and are unbiased.
		\item The image correlations can be decomposed into three groups, one (the first integral of equation~(\ref{eq:imagecorrelation2})) of which dominates if the conditions are right.
		\item That dominant peak is ONLY due to the self-correlation of individual particles with themselves, and only those particles present at the two instants of exposure.
		\item That the peak can be fitted with a curve (typically a 3-point Gaussian) to yield an estimate for the velocity at an instant, say $\vec{u}_{oi}(\vec{y},t)$. Note that velocity and displacement can be used interchangeably given our assumptions.
		\item The `velocity at an instant', $\vec{u}_{oi}(\vec{y},t)$, is a  quantized version of the velocity of the $n(\vec{y},t)$ particles in the volume at that instant, and each  of these contributions has an uncertainty of $P_i$ due to pixelation ($\pm 1/2$ pixel) plus any additional contributions including optical defects, etc.
		\item Since $n(\vec{y},t)$ is both finite and random, $\vec{u}_{oi}(\vec{y},t)$ is not the same as the instantaneous volume-averaged velocity.  This difference will prove to be the other primary noise source, and in many cases the dominant contribution.
	\end{enumerate}

In view of the above and since we are dealing only with the displacements (or velocities) of the image correlation peak, we (at least from the perspective of analysis) can forget about the images entirely.  And instead we treat the PIV output as simply an instantanous volume-average based only on the particles present in the interrogation volume at any instant.  Our $g$-functions provide an especially powerful tool for doing this, since we can avoid the messy problems of trying to do statistics on finite sums  with random elements.

\subsection{Representation of the PIV signal}

In practice the PIV produces an instantaneous velocity which is a
weighted average of the particles over the volume at any instant in time, the exact
weighting being determined by the contributions of the particles to
the correlation function produced by them.  We will assume at the
outset of this analysis that the time between flashes is
sufficiently short that we can assume that our individual particle
image displacements can indeed be interpreted as particle
velocities.  We will not, at least for the moment, consider sources
of error or noise that arise from practical problems of
implementation like hardware imperfections, quantization due to
discretization, determining the displacement peak of the correlation,
etc (\cite{foucaut04}). Rather we will focus only on the fluid
mechanics of the perfect instrument.
We will consider as fundamental, however, the intrinsic spatial
filtering due to the finite interrogation volume over which the PIV
must produce its `point' measurement, and as well the consequences
of the finite and random number of particles present in that
interrogation volume at any instant.

\subsection{The interrogation (or measuring) volume} 
Let $w(\vec{y},\vec{x}-\vec{y})$ be the interrogation volume (or measuring volume) centered at
location $\vec{y}$. (Note that this implicitly assumes that all measuring volumes are the same, which is usually the case in PIV except in case of interrogation window deformation.)  The integral of $w(\vec{y},\vec{r})$ over the (infinite)
domain defines the volume of the interrogation volume centered at location $\vec{y}$;
i.e.,

\begin{equation}
V(\vec{y}) \equiv \int_{as} w(\vec{y},\vec{x}-\vec{y}) d\vec{x}
\end{equation}
For PIV, a convenient choice for $w(\vec{y},\vec{x}-\vec{y})$ is the top-hat function given by:

\begin{eqnarray}
& & 1;  \hspace{0.04 in} |x_1 - y_1| \le \Delta_1/2,\hspace{0.1in} |x_2-y_2| \le \Delta_2/2,|x_3-y_3| \le \Delta_3/2  \nonumber \\
 w(\vec{y},\vec{x}-\vec{y})
 & = & \label{eq:volume}\\
& &   0; \hspace{0.04in} otherwise. \nonumber 
\end{eqnarray}
In this case $V = \Delta_1 \Delta_2 \Delta_3$.

\subsection{The particles in the interrogation volume}
The Lagrangian displacement field, $\vec{X}(\vec{a},t)$ gives the
present location of the particle which was initially ($t=t_o$) at
location $\vec{a}$; i.e.,

\begin{eqnarray}
\vec{x} & = & \vec{X}(\vec{a},t) \\
\vec{a}& = & \vec{X}(\vec{a},t_o)
\end{eqnarray}
where $t_o$ is the initial time (or instant of particle random insertion).
The primary reason for the Lagrangian representation is that it is
really only at the initial instant that we have any hope of being
able to say with confidence that the locations of the particles are
statistically independent from those of the velocity field. In
particular, by making this hypothesis we ensure that the statistics
of the particle locating functions, the $g(\vec{a})$, are
independent from those of the velocity or displacement fields.
Whether we can in fact make this happen as practical matter is
something else, but chances are probably greater than with any other
beginning hypothesis.

Since the $g(\vec{a})$'s themselves can be viewed as
$\delta$-functions which are located wherever there is an actual
scattering particle, it follows that the number of scattering
particles in the interrogation volume at any instant in time at
location $\vec{y}$, say ${n}(\vec{y},t)$ is given by:

\begin{equation}
{n}(\vec{y},t) =  \int_{as} w(\vec{y},\vec{X}[\vec{a},t]-\vec{y}) g(\vec{a}) d\vec{a} \label{eq:instantnumber}
\end{equation}
The $g$'s say whether there was a particle at location $\vec{a}$,
while the Lagrangian displacement field, $\vec{X}(\vec{a},t)$, fixes
the argument of $w(\vec{y},\vec{x}-\vec{y})$, which in turn indicates when it
enters and leaves the interrogation volume.

It might be argued that we could have equally well represented this
in Eulerian space by writing $g(\vec{x},t)$, for example. The reason
we do not is that it is probably never true that $g(\vec{x},t)$ is statistically
independent from the velocity or displacement fields.  The experimental evidence that
the particles cluster would seem to confirm our concern. The crucial requirement is that the {\it sampling} process and the {\it sampled} process must be statistically independent.  Simply being uncorrelated is not enough.

It is important to note that although the $g(\vec{a})$ are usually assumed to be $\delta$-functions, {\em this does not imply that they can only represent infinitesimal particles (or images)}.  By putting a random amplitude with the g-functions, finite images or particles can be also be associated with a particular point in time and space -- exactly what is accomplished by the fit to the correlation peak of the images.

\subsection{The statistics of $n(\vec{y},t)$}

 If we assume the particles to have been initially
statistically uniformly distributed and define $\langle g(\vec{a})
\rangle = \mu$, it is quite easy to show from
equation~\ref{eq:instantnumber} that $\mu$ is the expected number of
particles {\it per unit volume}. First compute the average number of
particles in the interrogation volume, say $N(\vec{y})$, to obtain:

\begin{eqnarray}
N(\vec{y}) = \langle n(\vec{y},t)\rangle & = & \int_{as} \langle w(\vec{y},\vec{X}[\vec{a},t]-\vec{y}
) g(\vec{a})\rangle  d\vec{a} \nonumber \\
& = & \mu \int_{as} \langle w(\vec{y}, \vec{X}[\vec{a},t] - \vec{y})
\rangle \hspace{0.02in}  d\vec{a} \label{eq:instantnumberavg}
\end{eqnarray}

Now we can map the last integral from Lagrangian to Eulerian
coordinates using $d\vec{a} = J^{-1} d\vec{x}$, where $J$ is the
ratio of volume elements given by the Jacobian
$J=|\partial(X_1[\vec{a},t],X_2[\vec{a},t],X_3[\vec{a},t])/\partial{(a_1,a_2,a_3)}|$.
For incompressible flow, $J=1$, so we obtain:

\begin{equation}
N(\vec{y}) = \mu \int_{as} w(\vec{y},\vec{x}-\vec{y}) ~ d\vec{x} = \mu V(\vec{y})  \label{eq:muVN}
\end{equation}

Note this equivalence of $N$ and $V$ for constant $\mu$ means that either can be used. Also for fixed third dimension (the sheet thickness) it will be easy to confuse dependence on $N$ with dependence on the interrogation area, $\Delta_1 \Delta_2$.  But it is really the average number of images in the interrogation volume, $N$, that is the fundamental independent variable.

\section{The instantaneous PIV output and the Eulerian volume-averaged velocity}

Since the quantity we associated with the displacement of the primary correlation peak was directly related to the correlation of particles with themselves and cross-correlations were eliminated,  we can now relate $\vec{u}_{oi}(\vec{y},t)$ directly to the velocities (or equivalently displacements) of individual particles in the volume and explore how it relates to volume-average Eulerian quantities we would like to obtain.

\subsection{The instantaneous PIV output}

  Since we are using only the image correlation peaks, we can define the concept of ${u}_{oi}(\vec{y},t)$ more precisely as:

\begin{equation}
{u}_{oi}(\vec{y},t) = \frac{1}{N(\vec{y})}\int_{as} {v_i}(\vec{a},t) w(\vec{y},\vec{X}[\vec{a},t]-\vec{y})
g(\vec{a}) d\vec{a} \label{eq:basicunbiased}
\end{equation}
where $N(\vec{y}) =\langle n(\vec{y},t) \rangle = \mu V(\vec{y})$, $V(\vec{y})$ is the volume size and ${v_i}(\vec{a},t)$ is the individual Lagrangian particle velocity.  

The reason for using the average number of images in the interrogation window\footnote{This is usually specified as particles per pixel or simply `ppp'.}, $1/N(\vec{y})$, instead of the instantaneous value, $1/n(\vec{y},t)$, is explained in Appendix A and B where the instantaneous alternative is shown to be biased.
Note that this finite-$N$-estimator, ${u}_{oi}(\vec{y},t)$, has been obtained by summing over the velocities of all the images (or particles) in the interrogation volume at an instant, exactly like the `velocity' (or `displacement') correlation peak analysis has presented us with.

In the remaining sections, unless there is ambiguity, we will drop the explicit dependence of $N$ and $V$ on $y$. But in practice it must always be considered since no two interrogation volumes need be alike.

\subsection{The instantaneous `volume-averaged' velocity}
The goal of any finite probe measurement is to obtain the {\it instantaneous volume-averaged velocity} defined by:

\begin{equation}
\tilde{u_i}(\vec{y},t) =  \frac{1}{V} \int_{as}
{u_i}(\vec{x},t) w(\vec{y},\vec{x}-\vec{y}) d\vec{x}
\label{eq:Eulerianinstantvolumeavgdvelocity}
\end{equation}
where ${u_i}(\vec{x},t)$ is the local Eulerian velocity.  Note
that this assumes that {\it all the fluid particles in the volume}
contribute equally to the integral.  Note that in the limit as $V \rightarrow
0$, $w(\vec{y},\vec{x}-\vec{y})/V \rightarrow \delta(\vec{x}-\vec{y})$, so the left and
right-hand sides produce exactly the same velocity as the volume
shrinks to a point.

What we would really like to be able to say is that
${u}_{oi}(\vec{y},t)$ is the same as the instantaneous
volume-averaged Eulerian velocity, $\tilde{u_i}(\vec{y},t)$.
In fact, our PIV instantaneous output, ${u}_{oi}(\vec{y},t)$, is at best an
approximation, since all fluid particles in the space are not
represented by our small sample of seeding particles, even if they
follow the flow. For example, all of the particles at one instant of
time might be in one corner of the measuring volume at the same time
that there is a significant velocity gradient across it.   We can
see this difference using our model if we map
equation~\ref{eq:Eulerianinstantvolumeavgdvelocity} into Lagrangian
coordinates; i.e.,

\begin{eqnarray}
\tilde{u_i}(\vec{y},t) & = & \frac{1}{V} \int_{as}
{u_i}(\vec{x},t) w(\vec{y},\vec{x}-\vec{y}) d\vec{x} \nonumber \\
&  = & \frac{1}{V}\int_{as} {v_i}(\vec{a},t)
w(\vec{y},\vec{X}[\vec{a},t]-\vec{y}) d\vec{a},
\label{eq:Lagrangianinstantvolumeavgdvelocity}
\end{eqnarray}
This can be compared to equation~(\ref{eq:basicunbiased}) from which the
obvious difference is the absence of
$g(\vec{a})$. The reason for the latter is, of course, that ALL of the
fluid particles contribute to the integral, not just the ones with
scattering particles.   Note that we have assumed the flow to be
incompressible so that the ratio of volume elements, the Jacobian of the
displacement transformation is unity; i.e., $J =
|\partial(X_1,X_2,X_3)/\partial(a_1,a_2,a_3)| = 1$.

Clearly the reasons our PIV output, equation~(\ref{eq:basicunbiased}), is not the same as equation~(\ref{eq:Eulerianinstantvolumeavgdvelocity}) (or
equation~(\ref{eq:Lagrangianinstantvolumeavgdvelocity})) and is only an
`approximation' are
two-fold: first the particles might not be following the flow; and
second, the limited number of particles in the interrogation volume
at any instant means we have at best a crude approximation to the
instantaneous Eulerian volume-averaged velocity.  {\it Exploring the consequences of the second is the primary concern here.  The reason is that
any departures from the true volume-averaged
instantaneous velocity must be interpreted as either a {\it bias}
(if the mean difference is non-zero) or {\it noise} (on the statistics of the
fluctuating quantities) on the data, albeit noise resulting from the
flow itself.}

Given the differences between equation~(\ref{eq:basicunbiased}) and
equation~(\ref{eq:Lagrangianinstantvolumeavgdvelocity}), especially
if there is a relatively small number of particles in any
interrogation window, the best we can hope for is that the {\it
statistics} of ${u}_{oi}(\vec{y},t)$ be the same as
$\tilde{u_i}(\vec{y},t)$. In order to investigate this
further, we will need to assume that the particles are following the
flow exactly. If the particles are {\it not} following the flow,
then the most we can say is that the statistics of
${u_o}_i(t)$ represent the statistics of a volume-average of
the velocity of the particles which happen to pass through the volume.
This is a sort of Eulerian-particle hybrid quantity -- the
statistics of particles as they pass a given point in space. The
multi-phase flow people probably have a name for this, since it
would occur all the time in their problems where the particles
almost never follow the flow.

\subsection{Quantization (or pixelization) noise \label{sec-pixelnoise}}
There is one big difference between PIV and many other instruments in that the discretization into pixels (or bits) is done BEFORE the integration over the volume is carried out.  In a hot-wire signal, for example, the averaging (or integration) over the wire-length is performed before the data is sampled by an A/D converter.  Thus the quantization noise is added  to the hot-wire signal AFTER the spatial filtering, so it is simply added as an unfiltered white noise.  

But in PIV the whole image correlation analysis is carried out digitally. So the spatial integral of equation~(\ref{eq:basicunbiased}) really should be represented this way:

\begin{equation}
{u}_{oi}(\vec{y},t) = \frac{1}{N}\int_{as} [{v_i}(\vec{a},t)~+~P_i(\vec{a},t)]~w(\vec{y},\vec{X}[\vec{a},t]-\vec{y})~
g(\vec{a}) d\vec{a} \label{eq:basicunbiased3}
\end{equation}
where the $P_i(\vec{a},t)$ include all optical and quantization noise sources, so must be at a minimum $\pm 1/2$-pixel (or its dimensional equivalent).

The importance of the above can not be underestimated, since it means that the effects of pixelization and the finite number of particles will {\em both} be spatially filtered.  And hence difficult to distinguish.  This appears to have first been noted by Lourenco and Krothapalli~\cite{LK2000}, and exploited extensively by ~\cite{foucaut04}.

In the following section we will focus first on the problem of only a limited number of samples within the volume as it affects the velocity estimate. We will defer the quantization noise discussion until Section~\ref{sec-pixel2}.

\section{Single point statistics of ${u}_{oi}(\vec{y},t)$}

Let's begin with a simple average (over an infinite ensemble) of
equation~(\ref{eq:basicunbiased}):

\begin{equation}
\langle {u}_{oi}(\vec{y},t) \rangle = \langle{
\frac{1}{N} \int_{as}
 {v_i}(\vec{a},t) w(\vec{y},\vec{X}[\vec{a},t]-\vec{y})
g(\vec{a})  d\vec{a} }\hspace{0.04in} \rangle\label{eq:mean1}
\end{equation}
In general we cannot assume the three quantities under the integral
on the right-hand-side to be statistically independent, since where
the particle is depends on its velocity history and that in turn on its
initial position.\footnote{This is a major difference from earlier analyses, some of which were not too careful on this point.}

\subsection{The importance of random initial particle positions}
 If, however, at $t=t_o$ we scatter the particles completely
randomly throughout the flow (our hypothesis), then we can be
positively sure that the initial position of the particles,
$\vec{a}$, is statistically independent from everything else. {\bf
In other words, by assuming the particles to be randomly distributed
{\it initially} we have succeeded in uncoupling the sampling process
from the flow, but only at $t=t_o$!}

So, let's try to proceed with equation~\ref{eq:mean1} directly {\it
without} transforming into Eulerian coordinates.  Because of the statistical independence in these coordinates, we can break the average into two parts; i.e.,

\begin{equation}
\langle {v_i}(\vec{a},t) w(\vec{X}[\vec{a},t]-\vec{y})
g(\vec{a})\rangle = \langle {v_i}(\vec{a},t)
w(\vec{y},\vec{X}[\vec{a},t]-\vec{y})\rangle \hspace{0.05in} \langle
g(\vec{a}) \rangle,
\end{equation}
This is true because we have ensured that the initial locations are
statistically independent, both of each other and of the velocity
field.  

Note that if we had included the factor of
$1/{n}(\vec{y},t)$ instead of $1/N$ we could not do this because $\langle 1/{n}(\vec{y},t) \rangle \ne 1/N$ in general.  Clearly
this factor is problematical, at least from the point-of-view of
statistical analysis. As shown in the Appendices, it is clear that care must be taken to ensure that we have not implicitly introduced it with our peak analysis.

\subsection{The mean value: $\langle {u}_{oi}(\vec{y},t \rangle)$ }

It follows immediately that:

\begin{eqnarray}
& &  \langle {u}_{oi}(\vec{y},t) \rangle  =   \frac{1}{N}
\int_{a s}
 \langle {v_i}(\vec{a},t) w(\vec{y},\vec{X}[\vec{a},t]-\vec{y})\rangle
 \hspace{0.05in} \langle g(\vec{a}) \rangle \hspace{0.05in}
 d\vec{a} \label{eq:meanvalueuoi}
\end{eqnarray}

We apply the same procedure as above, and map the
integral on the right-hand side to Eulerian space. Noting that $N =
\mu V$ it follows immediately that:
\begin{equation}
\langle {u_o}_i(\vec{y},t) \rangle  =  \frac{1}{V} \int_{as}
\langle {u_i}(\vec{x},t) w(\vec{y},\vec{x}-\vec{y}) \rangle d\vec{x}
\end{equation}
But this is just the average of the instantaneous volume-averaged
Eulerian velocity defined by
equation~(\ref{eq:Eulerianinstantvolumeavgdvelocity}); i.e.,

\begin{equation}
\langle {u}_{oi}(\vec{y},t) \rangle  =  \langle
\tilde{u_i}(\vec{y},t) \rangle \label{eq:mean2}
\end{equation}
which is exactly the result we had hoped for.  

But note that this unbiased mean has only been obtained by NOT using $1/n(t)$ in the estimator.  If we had we would have discovered an extra factor of $1+ 1/N$ which is clearly biased and vanishes slowly with increasing $N$.  This is discussed in detail in the appendices.

\section{Two-point statistics of ${u}_{oi}$.}

Now let's compute the two-point statistics of the velocity
fluctuation, $u_{oi}(\vec{y},t)$, defined by:

\begin{equation}
u_{oi}'(\vec{y},t) = {u}_{oi}(\vec{y},t) - \langle u_{oi}(\vec{y},t) \label{eq:flucvelPIV}
\rangle
\end{equation}
We learned above in equation~(\ref{eq:mean2}) that $\langle
u_{oi}(\vec{y},t) \rangle = \langle \tilde{u_i}(t) \rangle $. 
 What we would {\it like} to be true is that
$u_{oi}'(\vec{y},t)$ has the same {\it two-point}
statistics as the instantaneous fluctuating {\it volume-averaged}
Eulerian velocity field, $\tilde{u}_{i}'(\vec{y},t)$, given by:
\begin{equation}
\tilde{u}_i'(\vec{y},t) =  \tilde{u_i}(\vec{y},t) - \langle
\tilde{u_i}(\vec{y},t) \rangle. \label{eq:flucvolavg}
\end{equation}

It is clear from the outset, however, that we can only expect this
to be approximately true, since at any given instant the PIV samples
only at a few points within the volume, namely those with scattering
particles. Therefore we should also expect any differences from the
instantaneous volume-average Eulerian velocity to show up as
`noise'.  The source of this `turbulence noise' is analogous to
turbulence and mean gradient ambiguity noise which affects the many-particle continuous-mode LDA (tracker mode)(v.\ \cite{GeorgeLumley1973,Buchhaveetal1979}); namely the deviations between our instantaneous
volume average over the actual {\it scattering} particles present
and the average over all {\it fluid} particles in the measuring
volume at the same instant.

\subsection{The source of the `turbulence noise'}

While we could proceed directly, it is a bit more insightful if we
pose the problem another way. Start by taking the difference
between the instantaneous fluctuating `unbiased' PIV output
velocity, $u_{oi}'(\vec{y},t)$, and the instantaneous fluctuating
volume-averaged Eulerian velocity, $\tilde{u}_i'(\vec{y},t)$ given by
equations (\ref{eq:flucvelPIV}) and (\ref{eq:flucvolavg}) to obtain:

\begin{equation}
\Delta u_{oi}(\vec{y},t) = u_{oi}(\vec{y},t) - 
\tilde{u}_i(\vec{y},t)  \label{eq:Defndelta}
\end{equation}
Note that from the above $\langle \Delta u_{oi}(\vec{y},t) \rangle = 0$.    Ideally, we would hope to
find that the two point statistics of $\Delta u_{oi}(\vec{y},t)$ are
identically zero as well.

 It follows from
equations~(\ref{eq:basicunbiased})  and, (\ref{eq:Eulerianinstantvolumeavgdvelocity})
 that:

\begin{eqnarray}
\Delta u_{oi}(\vec{y},t) & = &  \frac{1}{N}\int_{as} {v_i}(\vec{a},t) w(\vec{y},\vec{X}[\vec{a},t] -\vec{y}) g(\vec{a}) d\vec{a}\nonumber \\&& - \frac{1}{V}
\int_{a s}
 {u_i}(\vec{x},t) w(\vec{y},\vec{x}-\vec{y})
 \hspace{0.05in}
 d\vec{x} \label{eq:velocdiff}
 \end{eqnarray}
where the second integral is $\tilde{u}_{i}(\vec{y},t)$.   Now we
transform the second integral into its equivalent Lagrangian frame
(assuming incompressibility) to obtain:

\begin{eqnarray}
 \Delta u_{oi}(\vec{y},t) & = &  \frac{1}{N}\int_{as} {v_i}(\vec{a},t) w(\vec{y},\vec{X}[\vec{a},t]-\vec{y}) g(\vec{a}) d\vec{a} \nonumber \\& & - \frac{\mu}{N}
\int_{a s}
 {v_i}(\vec{a},t) w(\vec{y},\vec{X}[\vec{a},t]-\vec{y})
 \hspace{0.05in}
 d\vec{a} \label{eq:velocdiff2}
 \end{eqnarray}
where we have replaced $V$ by $\mu N$ using equation~\ref{eq:muVN}.
Note that the difference between the first and second integrals is
that the second integral is over {\it all} fluid particles, while
the first integral is only over those fluid particles which have
been replaced by scattering particles.  

This can be simplified by combining the integrals to obtain:

\begin{eqnarray}
\Delta u_{oi}(\vec{y},t) & = &  \frac{1}{N}\int_{as} {v_i}(\vec{a},t) 
 w(\vec{y},\vec{X}[\vec{a},t]-\vec{y})~[g(\vec{a})- \mu]~ 
d\vec{a} \label{eq:velocdiff3}
\end{eqnarray}
Using this we can compute the two-point single-time velocity difference correlation $\langle  \Delta u_{oi}(\vec{y},t) \Delta u_{oj}(\vec{y'},t)
\rangle $ as:

\begin{eqnarray}
& & \langle  \Delta u_{oi}(\vec{y},t) \Delta u_{oj}(\vec{y'},t)
\rangle  =  \frac{1}{N^2}  \times \label{eq:prodint} \\
& & \int\int_{as} \langle \left[ v_i(\vec{a},t) v_j(\vec{a'},t)   w(\vec{y},\vec{X}-\vec{y}) 
w(\vec{y'},\vec{X'}-\vec{y}')  [g(\vec{a})-\mu] [g(\vec{a}')-\mu] \right] \rangle \hspace{0.03in}  d\vec{a} d\vec{a'} 
\nonumber
\end{eqnarray}
where $\vec{y}$ and $\vec{y}'$ are two separate locations of the interrogation volume and $\vec{X} = \vec{X}\left[\vec{a},t\right]$ and $\vec{X'} = \vec{X}\left[\vec{a'},t\right]$.

But since the $g$'s are statistically independent from each other and the particle displacements, thanks to our defining them in initial position coordinates and distributing them randomly initially, we can split the averages under the integral to obtain:

\begin{eqnarray}
& & \langle  \Delta u_{oi}(\vec{y},t) \Delta u_{oj}(\vec{y'},t)
\rangle  = \frac{1}{N^2} \times \nonumber \\
&& \int\int_{all~space} \langle  v_i(\vec{a},t) v_j(\vec{a'},t)   w(\vec{y},\vec{X}-\vec{y}) 
w(\vec{y}',\vec{X'}-\vec{y'}) \rangle \label{eq:prodint} \\
& & \hspace{2in} \times ~  \langle [g(\vec{a})-\mu] [g(\vec{a}')-\mu] \rangle \hspace{0.03in}  d\vec{a} d\vec{a'} 
\nonumber
\end{eqnarray}
Recall that statistical independence of the scattering particle initial positions implied that:
$\langle [g - \mu]~[g'-\mu] \rangle = \mu \delta(\vec{a}' - \vec{a})$ (equation~(\ref{eq:ggprime1})).  Application to the double integral of equation~(\ref{eq:prodint}) yields immediately:

\begin{eqnarray}
 & & \langle \ \Delta u_{oi}(\vec{y},t)  \Delta u_{oj}(\vec{y'},t)
 \rangle   \label{eq:correlationfinal1} \\
 & & =\frac{\mu}{N^2} \int_{as}   \langle v_i(\vec{a},t) v_j(\vec{a},t) 
~w(\vec{y},\vec{X}-\vec{y}) 
w(\vec{y}',\vec{X}-\vec{y'}) \rangle ~ \hspace{0.03in}  d\vec{a} \nonumber
\end{eqnarray}
This integral is purely a
consequence of the random sampling of the flow by the particles and is seen as noise. Note that the
window-functions, the $w(~)$'s, under the integral sign are evaluated for two interrogation
volume locations, one centered at $\vec{y}$, the other at
$\vec{y'}$, but the particle displacement field is the same for
both. 

{\em Clearly if there are two different non-overlapping interrogation volumes, then the noise contribution is zero, since a particle
can not be in two places at once.} For two-time quantities, this will need to be reconsidered. But all of the single time quantities we are the most
interested in (for now) are included (e.g., the second
moments, spatial correlations and spatial spectra).  Note also that the general practice has been to make the interrogation volumes as large as spatial filtering allows to decrease the noise by increasing $N$. But equation~\ref{eq:correlationfinal1} suggests that using two non-overlapping smaller volumes might be advantageous since the noise would be uncorrelated (e.g., using two $16 \times 16$ or two $16 \times 32$, instead of $32 \times 32$.).

\subsection{The two-point noise correlation \label{subsec-two-point}}

The role of this term is easier to see if we map from Lagrangian
space back to Eulerian space to obtain:

\begin{eqnarray}
 & &\langle \ \Delta u_{oi}(\vec{y},t)  \Delta u_{oj}(\vec{y'},t)
 \rangle   = \nonumber \\
& &  \frac{1}{N} 
  ~\left\{ \frac{1}{V}\int_{a s} \langle
  u_i(\vec{x}) u_j(\vec{x}) \rangle ~ [w(\vec{y},\vec{x}-\vec{y})
  w(\vec{y}',\vec{x}-\vec{y'})] \hspace{0.03in}  d\vec{x} \right\}
\end{eqnarray}
{\em But this is just a volume integral of the single-point Reynolds stress tensor, $\langle
u_i(\vec{x}) u_j(\vec{x}) \rangle$, weighted by the window function product.}
Hence the primary effect on the noise correlation is the diminishing product of the $w$'s as
one slides by the other.  

For top-hat window functions their product can be written as: 

\begin{equation}
w(\vec{y},\vec{y}-\vec{x}) w(\vec{y}',\vec{y}'-\vec{x}) = w(\vec{y},\vec{y}-\vec{x})~ W(\vec{y},\vec{y}'),
\end{equation} 
where $W(\vec{y},\vec{y}')$ is the familiar triangle window given by:

\begin{eqnarray}\label{eq:slidingvolume}
W(\vec{y},\vec{y}')&\approx & \left[1 - \frac{|y'_1-y_1|}{\Delta_1} \right ]\left[1
- \frac{|y'_2-y_2|}{\Delta_2} \right ]\left[1 -
\frac{|y'_3-y_3|}{\Delta_3} \right ]; \nonumber \\
 &  &  \hspace{1 mm}
|y'_1-y_1|\le \Delta_1,|y'_2-y_2|\le \Delta_2,|y'_3-y_3|\le \Delta_3   \\
& &  \hspace{1 mm} 0;  \hspace{1 mm}  otherwise
\nonumber
\end{eqnarray}
where $V = \Delta_1 \Delta_2 \Delta_3$ defines the interrogation
volume. 

Thus the two-point (single-time) correlation of the
volume-averaged velocities is multiplied directly by
equation~(\ref{eq:slidingvolume}); i.e.,

\begin{eqnarray}
&& \langle \ \Delta u_{oi}(\vec{y},t)  \Delta u_{oj}(\vec{y'},t)
\rangle \\
&& = \frac{1}{N} \langle u_i(\vec{y},t) u_j(\vec{y},t)
\rangle^{vol}\left[1 - \frac{|y'_1-y_1|}{\Delta_1} \right ]\left[1 -
\frac{|y'_2-y_2|}{\Delta_2} \right ]\left[1 -
\frac{|y'_3-y_3|}{\Delta_3} \right ]  \nonumber
\end{eqnarray}
for $|\vec{y'}-\vec{y}| \le \Delta_1,\Delta_2,\Delta_3$ and zero
otherwise. The pre-factor is the volume-averaged velocity
correlation tensor defined by:

\begin{equation}
 \langle u_i(\vec{y},t)
u_j(\vec{y},t) \rangle^{vol} =\frac{1}{V} \int_{a s} \langle
u_i(\vec{x})u_j(\vec{x}) \rangle w(\vec{y},\vec{y}-\vec{x})
 \hspace{0.03in}  d\vec{x}
\end{equation}

\subsection{The two-point two-time noise correlation \label{sec-twotimetwpointcorrelation}}

Often, as it is very short, we can ignore the time difference between flashes when computing the correlations.  But systematically now experimenters shift the interrogation volumes (a technique called ``Window shifting'') to maximize the number of useful particles and minimize the errors.   So, as we shall see below, the two-point, two-time noise correlation is also of interest.

This two-point, two-time noise correlation is given by:

\begin{eqnarray}
& &\langle \ \Delta u_{oi}(\vec{y},t)  \Delta u_{oj}(\vec{y'},t')
\rangle   =  \\
& & = \frac{\mu}{N^2} \int_{a s} \langle
v_i(\vec{a},t) 
v_j(\vec{a},t') ~ w(\vec{y},\vec{X}[\vec{a},t]-\vec{y})
w(\vec{y}',\vec{X}[\vec{a},t']-\vec{y'}) \rangle d\vec{a}
\nonumber 
\label{eq:A12t2s}
\end{eqnarray}
But the displacement field at time $t'$ can be related to that at time $t$ as follows:

\begin{equation}
\vec{X}_i(\vec{a},t') =  \vec{X}_i(\vec{a},t) ~ + \int_{t}^{t'} v_i(\vec{a},t'') dt''
\end{equation}
Thus the effect of changing $t$ is the same as changing $\vec{y}'$ by the amount $\delta \vec{y}$ where:

\begin{equation}
\delta \vec{y}_i = \vec{y}_i' - \vec{y}_i = \vec{X}_i(\vec{a},t') - \vec{X}_i(\vec{a},t) = \int_{t}^{t'} v_i(\vec{a},t'') dt''
\end{equation}

We can separate the Lagrangian velocity into its mean and fluctuating parts, $v_i(\vec{a},t) = V_i(\vec{a},t) + v_i'(\vec{a},t)$. We assume all the particles have the same mean velocity when they arrive at the interrogation volume and that it equals the Eulerian mean velocity, $U_i(\vec{y})$.   Using this in the displacement field and defining $t' = t + \delta t$ yields:

\begin{equation}
\delta y_i =    U_i \delta t +\int_{t}^{t+\delta t} v_i'(\vec{a},t'') dt''
\end{equation}
So the average displacement $\langle \delta{y}_i \rangle$ corresponding a given time difference $\delta t$ is

\begin{equation}
\langle \delta{y}_i \rangle  =    U_i ~ \delta t
\end{equation}

It follows immediately that the effect of a shift in time is to displace the value of $\vec{y}'$ in the window $W(\vec{y},\vec{y}')$; i.e.,

\begin{equation}
W(\vec{y},\vec{y} + \vec{U} ~\delta t) = 1 - U_\beta ~ \delta t/\Delta_\beta
\end{equation}
for $U_\beta  \delta t \le \Delta_\beta$ and zero otherwise.  Note that there is no sum on $\beta$.

Equation~(\ref{eq:A12t2s})  can be transformed back to Eulerian coordinates and evaluated at $(\vec{y} + \vec{y}')/2$ using $\vec{y}' = \vec{y} + \vec{U} \delta t$.  This  will be used in the succeeding sections to account for the window shift employed in the PIV analysis.

\subsection{The two-point correlation tensor, $B_{oi,j}(\vec{y},\vec{y'})$ }

\noindent The two-point (single-time) velocity cross-correlation
tensor of our PIV fluctuation velocity can be obtained by simply adding this noise contribution above to the volume-averaged two-point velocity correlation.  The result is: 

\begin{eqnarray}
B_{oi,j}(\vec{y},\vec{y'}) & = & \langle u'_{oi}(\vec{y},t)
u'_{oj}(\vec{y'},t) \rangle \nonumber \\ & = &  \langle
\tilde{u'}_{i}(\vec{y},t) \tilde{u'}_{j}(\vec{y'},t)\rangle  \label{eq:twopointcorrelation} \\
& & + \frac{1}{N} \langle u'_i(\vec{y},t) u'_j(\vec{y},t)
\rangle^{vol} W(\vec{y},\vec{y'}) \nonumber 
\end{eqnarray}
Recall that the first term has the convolution of the $w$'s already in it due to the spatial filtering. 
Also, note that in the first term the velocity components are evaluated at different positions, but at the same position in the second.
The first term on the right-hand-side is exactly what we wanted it
to be, the two-point cross-correlation of the volume-averaged Eulerian
velocities.  The second term on the right-hand-side is the noise
resulting from the fact that the velocity at any instant is based on
only a finite number of scattering particles.  

\subsection{The single-point Reynolds stresses}
If the PIV processing is done without window shifting, the single point Reynolds stresses follow immediately
from equation~(\ref{eq:twopointcorrelation}) evaluated at
$\vec{y}=\vec{y'}$; i.e.,

\begin{eqnarray}
\langle u'_{oi}(\vec{y},t) u'_{oj}(\vec{y},t) \rangle & = & \langle
\tilde{u'}_{i}(\vec{y},t) \tilde{u'}_{j}(\vec{y},t)\rangle  + \frac{1}{N} \langle u'_i(\vec{y},t) u'_j(\vec{y},t)
\rangle^{vol} \\
& \simeq & \langle \tilde{u'}_{i}(\vec{y},t)
\tilde{u'}_{j}(\vec{y},t)\rangle \left[{ 1 + \frac{1}{N} }\right]\label{eq:onepointReynoldsstresses}
\end{eqnarray}
Thus the net effect of the `turbulence' noise on the single point
Reynolds stress is to increase the volume-averaged values by a
factor of $1 + 1/N$.  This can be interpreted either as an additive
noise, or as a bias.

Now if the processing is performed with window shifting, things are a bit different. This means that the two interrogation windows overlap only partly and that $W(\vec{y},\vec{y'})$ has to be looked at carefully. Suppose the shift is $\vec{U}\delta t$ where $\vec{U}$ is the mean velocity and $\delta t$ is the time interval between the two laser pulses. Equation (\ref{eq:onepointReynoldsstresses}) then becomes:

\begin{eqnarray}
\langle u'_{oi}((\vec{y}+\vec{y'})/2,t) u'_{oj}((\vec{y}+\vec{y'})/2,t) \rangle & = & \langle
\tilde{u'}_{i}(\vec{y},t) \tilde{u'}_{j}(\vec{y},t)\rangle  + \frac{W(\vec{y},\vec{y'})}{N} \langle u'_i(\vec{y},t) u'_j(\vec{y},t)
\rangle^{vol} \\
& \simeq & \langle \tilde{u'}_{i}(\vec{y},t)
\tilde{u'}_{j}(\vec{y},t)\rangle \left[{ 1 + \frac{W(\vec{y},\vec{y'})}{N} }\right]\label{eq:Reynoldsstressesshift}
\end{eqnarray}
where $W(\vec{y},\vec{y}')$ is evaluated with $\vec{y}' = \vec{y} + \vec{U}~\delta t$. Note that if there is no shift, $W(\vec{y},\vec{y'}) = 1$ as $\vec{y} =\vec{y'}$, giving back equation (\ref{eq:onepointReynoldsstresses}).
 
Two things are interesting about this result.  First it can be
corrected if $N$ could be determined.  Second, the `noise' is
proportional to the mean square fluctuating velocity itself, not just the part of it which has
been filtered out.  This is quite different from the continuous and
burst-mode LDA where it is the fluctuatating gradients within the
volume which contribute to the noise; i.e., the part of the
turbulence that has been filtered by the finite volume.  Here the
result is determined instead by the uncertainty in the moments due
to the finite number of particles.

\subsection{Spectra}
If the three-dimensional velocity field is homogeneous (or even
locally homogeneous), then $\vec{y'} = \vec{y} + \vec{r}$ and $B_{oi,j}$ becomes a fuction of $\vec{r}$ only. We can easily perform the three-dimensional
Fourier transform of equation~(\ref{eq:twopointcorrelation}) to obtain the three-dimensional spectrum of the PIV output (c.f.\cite{george13}). The result is:

\begin{equation}
F_{oi,j}(\vec{k})  =  \left[\frac{1}{2 \pi} \right]^3 \int_{as}
e^{-i \vec{k}\cdot \vec{r}} B_{oi,j}(\vec{r}) d\vec{r}
\label{eq:specdefn}
\end{equation}
where we have suppressed the time dependence.  Note that there are
actually nine individual spectra since it is a tensor. Also note
that these are {\bf not} the one-dimensional spectra which will be
considered later.

It follows immediately by substituting
equation~(\ref{eq:twopointcorrelation}) into
equation~(\ref{eq:specdefn}) that:

\begin{equation}
F_{oi,j}(\vec{k}) = \tilde{F}_{i,j}(\vec{k}) + \frac{1}{N}\langle
u'_i(\vec{y},t) u'_j(\vec{y},t) \rangle^{vol} \hat{W}(\vec{k})
\label{eq:PIV3Dspectrum1}
\end{equation}
where $\hat{W}(\vec{k})$ is given by:

\begin{equation}
\hat{W}(\vec{k}) = \frac{\Delta_1 \Delta_2 \Delta_3}{(2 \pi)^3}
\left[\left(\frac{\sin k_1 \Delta_1/2}{ k_1
\Delta_1/2}\right)^2\left(\frac{\sin k_2 \Delta_2/2}{ k_2
\Delta_2/2}\right)^2 \left(\frac{\sin k_3 \Delta_3/2}{ k_3
\Delta_3/2}\right)^2\right]
\end{equation}
The first term is, of course, the three-dimensional spectrum of the
volume averaged velocity, which is in turn given by:

\begin{equation}
\tilde{F}_{i,j}(\vec{k}) = F_{i,j}(\vec{k}) \hat{W}(\vec{k})
\label{volspec}
\end{equation}
But the $\hat{W}(\vec{k})$ can be factored out of both terms leaving:

\begin{equation}
F_{oi,j}(\vec{k}) = \left\{ F_{i,j}(\vec{k}) + \frac{1}{N}\langle
u'_i(\vec{y},t) u'_j(\vec{y},t) \rangle^{vol} \right\}
\hat{W}(\vec{k}),
\end{equation}
so the net effect is as though a white noise were added to the
three-dimensional spectrum and their sum filtered through the
interrogation volume window.

In summary, the three-dimensional spectrum consists of two
contributions:  the spectrum of the volume-averaged velocities (or
the spatially filtered spectrum, if you will), and a noise
contribution which is proportional to the single point moments
divided by the expected number of particles in the interrogation
volume and multiplied by the sinc-function squared.  Clearly the
`noise' term can be minimized only by making the volume as small as
possible, while simultaneously making $N$ as large as possible -- a
nice trick if you can do it.

\subsection{The one-dimensional spectra}

The one-dimensional spectra (which are usually the ones of most
interest) can be computed two ways (c.f.\cite{george13}).  The first is by integrating the
three-dimensional spectrum over the two wavenumbers not
corresponding to the direction of interest.  For example, the
streamwise spectrum, say $F_{oi,j}^{1}(k_1)$, could be determined from:

\begin{equation}
F_{oi,j}^{1}(k_1) = \int\int_{-\infty}^\infty F_{oi,j}(\vec{k})
dk_2dk_3
\end{equation}
Alternatively it can be computed by evaluating the two-point
correlation in one direction only and then Fourier transforming it.
For example, the same spectrum as above could be determined from:

\begin{equation}
F_{oi,j}^{1}(k_1) = \frac{1}{2 \pi} \int_{-\infty}^\infty e^{-i
k_1 r} B_{oi,j}(r,0,0) dr
\end{equation}

In practice, of course, these  are usually computed over finite
domains using a finite spectral estimator and an FFT. While this
will add an additional window into the mix, it does not affect our
theoretical result here.  The easiest to consider theoretically is
the second method.  Let's use it to deduce the streamwise spectrum
(i.e., direction of separation same as velocity components).

First, evaluate the two-point correlation of
equation~(\ref{eq:twopointcorrelation}) for $i=1,j=1$ and $\vec{r} =
\vec{y'} - \vec{y} = (r,0,0)$ to obtain:

\begin{equation}
B_{o1,1}(r,0,0) = \tilde{B}_{1,1}(r,0,0)  + \frac{1}{N} \langle u'_1u'_1 \rangle^{vol} W(r,0,0) \label{eq:1-1corr}
\end{equation}
where $W_{1,1}(r,0,0) = 1 - |r|/\Delta$ for $|r| \le \Delta$ and zero
otherwise.  Note that $B_{1,1}(r,0,0) = \langle \tilde{u}_1(x,y,z)
\tilde{u}_1(x+r,y,z) \rangle$ is just the two-point correlation of the
volume-averaged (or filtered) 1-1 velocity with separation in the
1-direction (i.e., the correlation of the PIV-output velocities).
The factor $\langle u_1 u_1 \rangle^{vol}$ is the volume-averaged
mean square fluctuating velocity, which is  the same as $\langle u_1^{vol} u_1^{vol}) \rangle$ for a tophat.

Second, perform the one-dimensional Fourier transform to obtain
$F_{o1,1}^{1}(k_1)$ as:

\begin{eqnarray}
F_{o1,1}^{1}(k_1) & = & \frac{1}{2 \pi} \int_{-\infty}^\infty
e^{-ik_1r} B_{o1,1}(r,0,0) dr  \\
& = & \tilde{F}_{1,1}^{1}(k_1) + \frac{1}{N} \langle {u'_1}^2 \rangle^{vol}
\frac{\Delta_1}{2\pi} \left[\frac{sin (k_1 \Delta_1/2)}{k_1
\Delta_1/2}\right]^2  \label{eq:1Dspectrum}
\end{eqnarray}
where $\tilde{F}_{1,1}^{1}(k_1)$ is the one-dimensional spectrum of the
volume-averaged Eulerian velocity (our desired result) given by:

\begin{equation}
\tilde{F}_{1,1}^{1}(k_1) = \int\int_{-\infty}^\infty F_{1,1}(\vec{k})
\hat{W}(\vec{k}) dk_2 dk_3 \label{eq:1Dspectrum2} 
\end{equation}
Note that in equation~\ref{eq:1Dspectrum2} the $sinc^2(k_1 \Delta_1/2)$ part of $\hat{W}(\vec{k})$ can be factored outside the double integral.  {\em  So this is exactly the form `assumed' by Foucaut et al.\ ~\cite{foucaut04}, and utilized extensively in subsequent publications for optimization.}
	
A similar result can be obtained for any other one-dimensional
spectra of interest, including other directions and velocity
components, just by perturbing the indices. Thus, like the
corresponding three-dimensional spectrum, the one-dimensional
spectra consist of two parts: first the part we expected and wanted
corresponding to the spectrum of the volume-averaged velocity; and a
second `turbulence' noise spectrum equal to the variance multiplied
by a sinc-squared function and divided by the expected number of
particles in the interrogation volume.

\section{Effect of pixelization noise \label{sec-pixel2}}

It was noted in Section~\ref{sec-pixelnoise} that the discretization errors appear inside the defining integral, and are not simply additive to it; i.e., from equation~(\ref{eq:basicunbiased}),
	
	\begin{equation}
	{u}_{oi}(\vec{y},t) = \frac{1}{N}\int_{as} [{v_i}(\vec{a},t)~+~P_i(\vec{a},t)]~w(\vec{y},\vec{X}[\vec{a},t]-\vec{y})
	g(\vec{a}) d\vec{a} 
	\end{equation}
	where $P_i(\vec{a},t)$ represents the quantization noise source.
	Carrying out all of the same steps performed above for velocities leads to a very similar result for the pixel noise contribution to the two-point correlation.  Instead of equation~(\ref{eq:correlationfinal1}) we have:
	\begin{eqnarray}
	\langle \ \Delta u_{oi}(\vec{y},t)  \Delta u_{oi}(\vec{y'},t)
	\rangle & = & \frac{1}{N}  
 \frac{1}{V} \int_{as}   \langle (v_i(\vec{a},t)  + P_i(\vec{a},t))~(v_j(\vec{a},t) + P_j(\vec{a},t) ) \nonumber \\ & &
\times~  [w(\vec{y},\vec{X}[\vec{a},t]-\vec{y}) 
	w(\vec{y'},\vec{X}[\vec{a},t]-\vec{y'})  \rangle  d\vec{a} 
	\label{eq:correlationfinal2a}
	\end{eqnarray}
But the $P_i$ are uncorrelated with either the velocities {\em or  each other}.  So this can be reduced to:

\begin{eqnarray}
\langle \ \Delta u_{oi}(\vec{y},t)  \Delta u_{oi}(\vec{y'},t)
\rangle & = & \frac{1}{N} 
 \frac{1}{V} \int_{as}   \langle (v_i(\vec{a},t) v_j(\vec{a},t) \rangle  +  \langle P_m(\vec{a},t)~P_m(\vec{a},t)  \delta_{ij} )  \nonumber
  \\
& & ~~~~~ \times  ~w(\vec{y},\vec{X}[\vec{a},t]-\vec{y}) 
w(\vec{y}',\vec{X}[\vec{a},t]-\vec{y'}) \rangle  d\vec{a}  \label{eq:correlationfinal2}
\end{eqnarray}
If the $P_i$ are just a standard quantization with a uniform probability of all values between $\pm \Delta/2$ where $\Delta$ is the pixel dimension expressed in the same units as the velocity, then
$\langle P_m P_m \rangle = \Delta^2 / 12$. So it simply adds to the normal stresses components of the two-point correlation; i.e.,
equation~(\ref{eq:twopointcorrelation}) becomes:
	
	\begin{eqnarray}
	B_{oi,j}(\vec{y},\vec{y'}) & = & \langle u'_{oi}(\vec{y},t)\rangle
	u'_{oj}(\vec{y'},t) \nonumber \\ & = &  \langle
	\tilde{u'}_{i}(\vec{y},t) \tilde{u'}_{j}(\vec{y'},t)\rangle  \label{eq:twopointcorrelation2} \\
	& & + \frac{1}{N} \left\{ \langle u'_i(\vec{y},t) u'_j(\vec{y},t)
	\rangle^{vol} + \frac{1}{12}\Delta^2 \delta_{ij} \right\}   W(\vec{y},\vec{y'})\nonumber
	\end{eqnarray}
{\bf Note that this is very different from the usual quantization of continuous signals, since the usual quantization  variance has been reduced by a factor of $1/N$ and has the same correlation shape as the window product.}   Additional contributions to $P_i$ can be taken into account by simply adding their variance $\langle \eta_i \eta_j \rangle$ divided by N to equation (\ref{eq:twopointcorrelation2}).  It should be also noted that it is the $1/N$ multiplying the $\Delta^2/12$ that has previously been interpreted as ``sub-pixelization'.  

The error contribution to the each component of the {\em velocity variance} is easily computed by setting $i = j$ and evaluating equation~\ref{eq:twopointcorrelation2} at $(\vec{y} + \vec{y'})/2$. Using $W(\vec{y},\vec{y} + \vec{U} \delta t)$ if there is a window shift yields:

\begin{equation}
 \langle \Delta_{oi}^2 \rangle =\frac{1}{N} \left\{ \langle u'_i(\vec{y},t) u'_i(\vec{y},t)
\rangle^{vol} + \frac{1}{12}\Delta^2  \right\}W(\vec{y},\vec{y} + \vec{U} \delta t) \label{eq:intensityerror}
\end{equation}
(no sum on $i$).

The {\em three-dimensional spectrum} of equation~(\ref{eq:PIV3Dspectrum1}) becomes:
\begin{equation}
F_{oi,j}(\vec{k}) = \tilde{F}_{i,j}(\vec{k}) + \frac{1}{N} \left\{ \langle
u'_i(\vec{y},t) u'_j(\vec{y},t) \rangle^{vol} + \frac{\Delta^2}{12} \delta_{ij} \right\} \hat{W}(\vec{k})
\label{eq:PIV3Dspectrum2}
\end{equation}
And the corresponding {\em one-dimensional spectrum} of equation~(\ref{eq:1Dspectrum}) becomes:

\begin{eqnarray}
F_{o1,1}^{1}(k_1) = \tilde{F}_{1,1}^{1}(k_1) + \frac{1}{N} \left\{ \langle {u'_1}^2 \rangle^{vol}  +  \frac{\Delta^2}{12}  \right\}
\frac{\Delta_1}{2\pi} \left[\frac{sin (k_1 \Delta_1/2)}{k_1
	\Delta_1/2}\right]^2 \label{eq:1Dspectrumplusnoise}
\end{eqnarray}
We can define the {\em one-dimensional noise spectrum} to be:

\begin{equation}
F^{1}_{noise}(k_1) = \frac{1}{N} \left\{ \langle {u'_1}^2 \rangle^{vol}  +  \frac{\Delta^2}{12}  \right\}
\frac{\Delta_1}{2\pi} \left[\frac{sin (k_1 \Delta_1/2)}{k_1
	\Delta_1/2}\right]^2, \label{eq:1Dnoisespectrum3}
\end{equation}
which is exactly the form proposed and used extensively by Foucaut et al.\ \cite{foucaut04}, but with the amplitude  (bracketted term divided by $N$) determined empirically.


\section{Velocity differences and derivatives \label{sec-deriv}}

Of great interest is the ability of PIV to measure velocity derivatives.  There are various way this can be accomplished (see ~\cite{foucaut02}), but all ultimately involve differences of the velocity at two different locations, say $\vec{y}$ and $\vec{y}'$.  From equation~(\ref{eq:basicunbiased}) it follows immediately that the instantaneous velocity difference is given by:
	\begin{eqnarray}
	&& {\delta}_i(\vec{y}',\vec{y})  = {u}_{oi}(\vec{y}',t)  - {u}_{oi}(\vec{y},t)
	\end{eqnarray}
It is easiest to work with the following form (from equation~\ref{eq:Defndelta}):

	\begin{equation}
	 u_{oi}(\vec{y},t) =  \tilde{u}_i(\vec{y},t)  + \Delta u_{oi}(\vec{y},t) 
	 \label{eq:Defndelta2}
	\end{equation}
	since ideally it is the differences (and derivatives) of the volume-averaged velocities that we seek, and any deviation from this is noise or bias.
	
	Substitution yields:
	
	\begin{eqnarray}
	&& {\delta}_i(\vec{y'},\vec{y})  = [\tilde{u}_i(\vec{y'},t)  - \tilde{u}_i(\vec{y},t)]   + \{ \Delta u_{oi}(\vec{y'},t) - \Delta u_{oi}(\vec{y},t) \}
	\end{eqnarray}
	The first term on the right-hand side in square bracket is the velocity difference we seek, the volume-averaged velocity difference; and the second-term in curly brackets is the noise.
	Note that we know from our previous analysis that the average of the curly-bracketed term is zero, since the average of both parts is zero. 
	
	Since it is the instantaneous noise contribution to $i$-th velocity difference that is of primary interest here, let's define this as:
	
	\begin{equation}
	\delta_i^{noise}(\vec{y'},\vec{y}) =  \{ \Delta u_{oi}(\vec{y'},t) - \Delta u_{oi}(\vec{y},t) \}
	\end{equation}
	
	Now we are mostly interested in second-moments of velocity derivatives, say $\epsilon_{ik} =\langle \delta_i \delta_k \rangle$ for which the averaged noise contribution is:
	
	\begin{eqnarray}
	\xi_{ik}& = &\langle \delta_i^{noise}(\vec{y'},\vec{y}) ~\delta_k^{noise}(\vec{y'},\vec{y}) \rangle  \nonumber \\
	& = & \langle \{ \Delta u_{io}(\vec{y'},t) - \Delta u_{io}(\vec{y},t) \} \{ \Delta u_{ko}(\vec{y'},t) - \Delta u_{ko}(\vec{y},t) \} \rangle \label{eq:epsilonik} \\
	& = &\langle \Delta u_{io}(\vec{y'},t)~\Delta u_{ko}(\vec{y'},t)  \rangle + \langle \Delta u_{io}(\vec{y},t)~\Delta u_{ko}(\vec{y},t) \rangle \\
	& & - \langle  \Delta u_{io}(\vec{y},t) ~\Delta u_{ko}(\vec{y}',t) -  \langle  \Delta u_{io}(\vec{y'},t) ~\Delta u_{ko}(\vec{y},t) \rangle \rangle\nonumber
	\end{eqnarray}
But these are all moments and correlation we computed in the preceding sections.  The last two terms are zero if there is no overlap of the interrogation volumes.  And the first two terms are just twice equation~(\ref{eq:intensityerror}); i.e.,

 \begin{equation}
 \xi_{ik} =\frac{2}{N} \left\{ \langle u'_i(\vec{y},t) u'_k(\vec{y},t)
 \rangle^{vol} + \frac{1}{12}\Delta^2 \delta_{ik} \right\}. \label{eq:differenceerror}
 \end{equation}
 As can be seen, the noise on the derivative moments is twice the noise on the velocity moments themselves.

%
%

\section{Comparison with Experiment}

To validate the proposed theory, it is necessary to have experimental results where the noise is carefully quantified. The extensive experiment by \cite{foucaut04} provides in fact all of the data needed to test the theory summarized above. They performed measurements in a fluid at rest to characterize the noise alone and then measurements in a turbulent boundary layer with different magnifications and different dynamic ranges. In their analysis based on spectra, they anticipated almost all of the results obtained here, and their results can be readily interpreted with the present theory as shown below.  They missed only the explicit relations among the number of images, the pixelization and the turbulence intensity.  Even so they noted the following:

\begin{quotation}
	 `` The noise also decreases
	as the number of particle images in the window, given by $XYN_p$, increases '' (\cite{foucaut04})
\end{quotation}
Since their $X,Y$ are our $\Delta_1,\Delta_2$, the planar dimensions of the interrogation volume, and $N_p$ is the number of particle images per pixel, their $XYN_p$ is exactly our $N$; i.e.,

\begin{equation}
N = XY N_p = \Delta_1 ~ \Delta_2~ N_p
\label{eq:NvsNp}
\end{equation}

\subsection{Relation of present theory with the results of Foucaut et al.\ \cite{foucaut04}}

Before looking at the data, it is worth noting the key equation (10) of their paper:

\begin{equation}
E_{11PIV} = \left(E_{11HWA} + E_{noise}\right)\left(\frac{sin ~kX/2}{kX/2}\right)^2
\label{spectrumfoucaut}
\end{equation}
where $E_{11PIV}$ is the one-dimensional spectrum measured by PIV, $E_{11HWA}$ is the spectrum measured by Hot Wire Anemometry at the same location (considered as the best approximation of the exact spectrum $F_{1,1}^1(k_1)$), $E_{noise}$ is the noise spectrum amplitude and  $X$ corresponds in our notation to $\Delta_1$. It should be noted that implicit in the experimental results is the assumption that the PIV spectrum without noise is the same as the HW spectrum.  As evidenced by equations \ref{volspec} and \ref{eq:1Dspectrum2}, this cannot be exactly true since the spatial filtering for the HW and PIV is different.  \cite{foucaut04} were able to account for the PIV streamwise filtering in their model with the $sinc^2$ term shown above,  but not  in the cross-stream directions.  Nonetheless, as the result below will illustrate, their estimates correspond nicely to the theory proposed in this paper.

Comparing their equation (\ref{spectrumfoucaut}) to our equation (\ref{eq:1Dspectrumplusnoise}), $E_{noise}$ can be expressed
as a function of our two noise contributions:

\begin{equation}
E_{noise} = \frac{1}{N}\left(\langle {u'_1}^2\rangle^{vol} + \frac{\Delta ^2}{12}\right)\frac{\Delta _1}{2\pi}
\label{enoise}
\end{equation}
Also their $\zeta$-parameter can be expressed as:

\begin{equation}
\zeta = \frac{1}{2\pi N_p}\left(\langle {u'_1}^2\rangle^{vol} + \frac{\Delta ^2}{12}\right) = E_{noise}.\Delta_2
\label{zeta}
\end{equation}
Note that $\zeta$ was introduced by \cite{foucaut04} in their equation (6) as a constant characterizing the noise in the PIV records.

Thus, the present theory relates the noise in the PIV records to the velocity fluctuations inside the interrogation window and to the quantization noise on the particle images. This noise can be reduced by increasing $N_p$ (or $N$), the number of particles in the interrogation window. If $\Delta$ is known and the turbulence intensity can be guessed (or known), the constant $\zeta$ need not be optimized using the data anymore, it is given by the theory.

Looking now at the contribution of \cite{foucaut04}, the results of four different PIV experiments are provided: An experiment in a fluid at rest, one in a turbulent flow with low magnification, the same experiment with high magnification and finally the same experiment with a high dynamic range. The first three experiments were performed with one of the very first PIV video cameras, the Pulnix TM9701 with a framing rate of 30 im/s, while the last experiment used a photographic camera, the Kodak DCS 460, with a framing rate of the order of seconds. Table \ref{table-param} gives the parameters of interest for the different experiments. Most parameters are explicit, $M$ is the magnification.

\begin{table}
	\caption{Parameters of the four experiments in \cite{foucaut04}.}
	\label{table-param}       
	\begin{center}
		\begin{tabular}{lllllll}
			\hline
		 & Camera &  Size  & Pixel size & bits  & f&f\# \\ 
		 &&Pixels&$\mu m ^2$&&mm&\\
			\hline
  Fluid at rest&Pulnix TM9701&768x484&11.6x13.6&8&50&8\\
  Small dyn. range&&&&&&\\
  Low mag.&Pulnix TM9701&768x484&11.6x13.6&8&50&8\\
  High mag.&Pulnix TM9701&768x484&11.6x13.6&8&50&16\\
  High dyn.range&DCS 460 &3072x2048&9x9&12&100&2.8\\
			\hline
			\hline
 &M& \multicolumn{2}{l}{Part. im. size}&$N_p$ & Dyn. range\\ 
&&x Pixels&y Pixels&Part./pix&Pixels\\
\hline
Fluid at rest&0.11&2.5-3&1.5-2&0.02&0\\
Small dyn. range&&&&&\\
Low mag.&0.11&2.5-3&1.5-2&0.04&3\\
High mag.&0.32&2.5-3&1.5-2&0.025&5.8\\
High dyn.range&0.173&1.6-2&1.6-2&0.025&15\\
\hline
		\end{tabular}
	\end{center}
\end{table}

\subsection{No-flow data}

Looking first at the experiment in a fluid at rest, its main interest is to characterise the quantization noise when $\langle {u'_1}^2\rangle^{vol} = 0$ in equations (\ref{enoise}) and (\ref{zeta}). Figure~\ref{fig:JMEta2004spectranoflow} from \cite{foucaut04}  gives a plot of the one-dimensional spectrum in the two directions with a 32x32 Interrogation Window (IW) and different overlaps of this IW. It clearly illustrates the $sinc^2$ behaviour of the noise spectrum as given by equation (\ref{eq:1Dnoisespectrum3}) or by equation (2) in \cite{foucaut04}.

\begin{figure}
	\centering 
	\resizebox{0.95\linewidth}{!}{\includegraphics[scale=1]{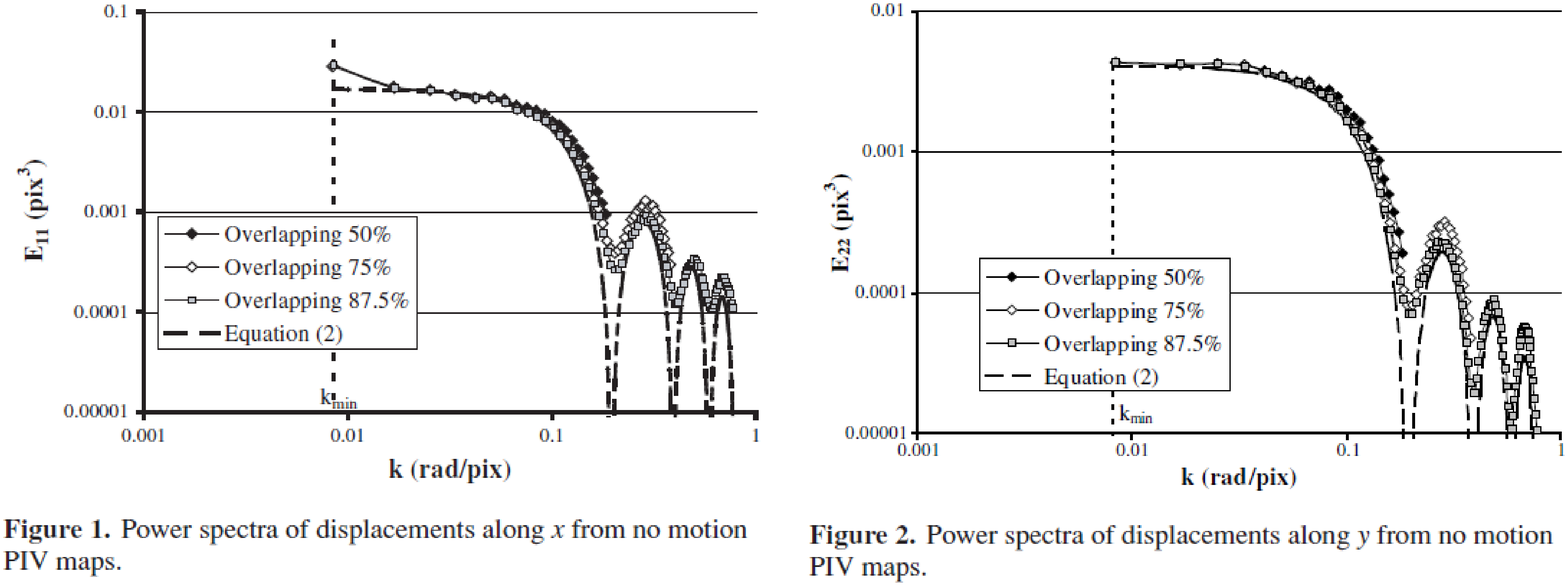}}
	\caption{Spectral data from no flow experiment from  \cite{foucaut04}. Equation (2) is in these authors paper and corresponds to equation (\ref{spectrumfoucaut}) here with no HWA spectrum.}
	\label{fig:JMEta2004spectranoflow}      
\end{figure}

Based on the fact that $\langle {u'_1}^2\rangle^{vol} = 0$, equation (\ref{zeta}) reduces to:

\begin{equation}
\zeta = \frac{1}{2\pi N_p} \frac{\Delta ^2}{12}
\label{zetanf}
\end{equation}
Also, since the fluid is at rest, there is no window shift so the variance of the noise is given by equation (\ref{eq:intensityerror}) for $i = 1$, $\langle {u'_1}^2\rangle^{vol} = 0$ and $W(\vec{y},\vec{y'}) = 1$; i.e.,

\begin{equation}
\sigma_{th} = \sqrt{\frac{1}{N}\frac{\Delta ^2}{12}}
\label{sigma}
\end{equation}

As mentionned earlier, our best estimation of the quantization error $\Delta$ is $\pm 1/2 px$. Using this value of 0.5 in equation (\ref{sigma}) gives an estimation $\sigma _{th}$ which compares quite favourably with the data of \cite{foucaut04} along $x_2$ but not along $x_1$. This can be explained by the fact that the Pulnix camera was one of the very first video cameras usable for PIV. Although it was not working any more with interline transfer, it was not yet a full frame transfer camera. In the progressive scan used, lines are transfered one after the other but the pixels still have to be moved all along the line. This explains the higher level of noise evidenced in the experiment. To take this specificity into account, as mentionned in section \ref{sec-pixel2} about equation (\ref{eq:twopointcorrelation2}), it is possible to add an extra noise parameter say $\langle \eta ^2 \rangle$ to equation (\ref{zetanf}) which reads now:

\begin{equation}
\zeta = \frac{1}{2\pi N_p} \left(\frac{\Delta ^2}{12} + \langle \eta ^2 \rangle \right)
\label{zetaetanf}
\end{equation}
and equation (\ref{sigma}) becomes:

\begin{equation}
\sigma_{th} = \sqrt{\frac{1}{N}\left(\frac{\Delta ^2}{12} + \langle \eta ^2 \rangle \right)}
\label{sigmaeta}
\end{equation}

\begin{table*}[h]
	\caption{Noise characteristics of the \cite{foucaut04} experiment in a fluid at rest}
	\label{noflow}       
	\begin{center}
		\begin{tabular}{lllllll}
			\hline
			IW & Pixels &  N  & $\zeta_1$ & $\zeta_2$  & $\zeta_{1th}$ & $\zeta_{2th}$ \\
			\hline
			$16 \times 16$ & 256&5.12 & 0.49 & 0.12 & 0.48 & 0.17  \\
			$32 \times 32$ & 1024&20.48 & 0.55 & 0.14 & 0.48 & 0.17 \\
			$64 \times 64$ & 4096&81.92 & 0.59 & 0.14 & 0.48 & 0.17 \\
			\hline
			IW & $\sigma_1$ & $\sigma_{1e}$  & $\sigma_{1th}$ & $\sigma_2$ & $\sigma_{2e}$  & $\sigma_{2th}$ \\
			\hline
	$16 \times 16$ &	0.087&	0.107&	0.109&	0.050&	0.053&	0.064\\
	$32 \times 32$&	0.055&	0.057&	0.055&	0.029&	0.028&	0.032\\
	$64 \times 64$ &	0.033&	0.029&	0.027&	0.015&	0.014&	0.016\\
			\hline
		\end{tabular}
	\end{center}
\end{table*}
\noindent Adjusting $\langle {\eta_1} ^2 \rangle = 0.04$ (while keeping $\langle {\eta_2} ^2 \rangle = 0.0$) gives the results gathered in table \ref{noflow} where IW is the interrogation window size, Pixels is the number of pixels in the interrogation window, N is the average number of particles in the interrogation window based on $N_p$ in Table \ref{table-param}, $\zeta_1$ and $\zeta_2$ are the noise parameter of \cite{foucaut04} along the two directions, $\zeta_{1th}$ and $\zeta_{2th}$ are  given by equation (\ref{zetaetanf}), $\sigma_1$ and $\sigma_2$ are the noise variances directly measured by \cite{foucaut04}, $\sigma_{1e}$ and $\sigma_{2e}$ are the noise variances estimated by \cite{foucaut04} using their equation (7) and $\sigma_{1th}$ and $\sigma_{2th}$ are the noise variances given by equation (\ref{sigmaeta}). All data except N are in pixels.

As can be seen, the present theory is in quite good agreement  with the experiment for both $\sigma_1$ and $\sigma_2$ (the prediction of $\sigma_2$ can be nearly perfect using $\Delta = 0.45$ instead of $0.5$). The extra noise along $x_1$ due to the transfer mode of the Pulnix camera can be taken into account favourably and the noise parameter $\zeta$ is also correctly predicted. This value of the noise parameter $\langle {\eta_1} ^2 \rangle$ will be kept in the following for the experiments involving the Pulnix camera.

\subsection{Turbulent boundary layer data}

Looking now at the experiments with flow, all three were performed in a turbulent boundary layer in a plane parallel to the wall at a wall distance of $30$ wall units.  Consequently $x_1$ is along the flow and $x_2$ is transverse. As HWA data where available only for $u_1$, the analysis of the one-dimensional spectrum and of the noise were performed only along $x_1$. Figure \ref{spectraflow} gives the one-dimensional spectra with a 32x32 interrogation window compared to the HWA one and to the model of equation (\ref{spectrumfoucaut}).

\begin{figure}
	\centering 
	\resizebox{0.95\linewidth}{!}{\includegraphics[scale=1]{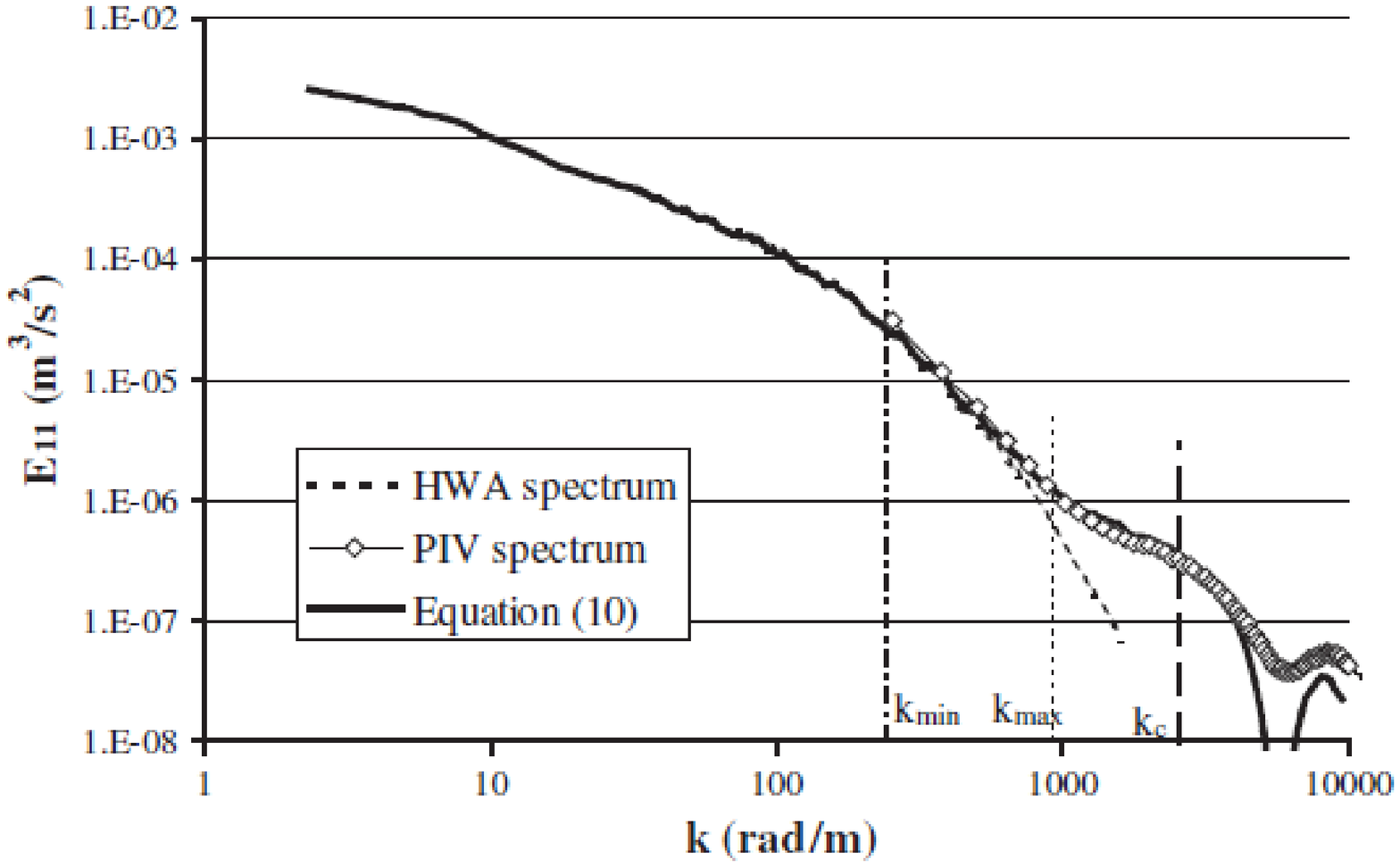}}
	\caption{Power spectra of velocity along $x_1$, 32 x 32 interrogation	window, high magnification. From  \cite{foucaut04} figure 14. Equation (10) is in these authors paper and corresponds to equation (\ref{spectrumfoucaut}) here.}
	\label{spectraflow}      
\end{figure}

This allowed the authors to determine $E_{noise}$ and $\zeta$ for the different cases. For these flow cases, window shifting was used by the authors to perform the PIV analysis. The noise variance can be estimated thanks to equation (\ref{eq:intensityerror}) and by taking into account the extra camera noise term $\langle \eta ^2 \rangle$:

\begin{equation}
\sigma_{th} = \sqrt{\frac{1}{N}\left(\langle {u'_1}^2\rangle^{vol} + \frac{\Delta ^2}{12} + \langle \eta ^2 \rangle\right)*W_1(U\delta t)}
\label{sigmafull}
\end{equation}
where $W(\vec{y},\vec{y}+\vec{U}\delta t)$ has been simplified to $W_1(U\delta t) = 1 - DR/\Delta_1$, where  $DR$ is the dynamic range given in table \ref{table-param} and the mean velocity  is in the 1-direction and given in pixels.

With such a window shifting, the relation between $\zeta$ (or $E_{noise}$) and $\sigma$ goes through the integration of the spectrum as detailed by \cite{foucaut04}, leading to their equation (7):

\begin{equation}
\sigma = \sqrt{\frac{4\zeta I}{\Delta_1\Delta_2}}
\label{sigmasp}
\end{equation}
where $I = 1.492$ is the integral over $[0,2\pi]$ of the sinc function. This equation can be inverted to give:
 
\begin{equation}
\zeta _{th} = \frac{\sigma ^2\Delta_1\Delta_2}{4 I}
\label{zetasp}
\end{equation}

Tables \ref{withflow1} to \ref{withflow3} provide both the data of Table 3 of \cite{foucaut04} and the present predictions with the same notations as in table \ref{noflow}. $N_p$ is the  number of particle per pixel (ppp) provided by \cite{foucaut04}. Here again, all data except N are in pixels. The ratio between the theoretical and experimental values is provided for both $\sigma$ and $\zeta$.

\begin{table*}[h]
	\caption{Noise characteristics of the \cite{foucaut04} experiments in a turbulent flow: Low magnification}
	\label{withflow1}       
	\begin{center}
		\begin{tabular}{lllllll}
			\hline
			$N_p$ =&0.04&&&&&\\
IW&	Pixels&	N& $W_1$&	$\sigma_1$&	$\sigma_{1th}$& Ratio\\
\hline
$16 \times 16$ &576&	23.04&0.875&	0.159&	0.100&0.629\\
$32 \times 32$ &1024&	40.96&0.906&	0.119&	0.076&0.641\\
$36 \times 36$&1296&	51.84&0.917&	0.106&	0.068&0.644\\
$64 \times 64$&4096&	163.84&0.953&	0.060&	0.039&0.652\\
\hline
IW&	$u'_1$&	${u'_1}^2$&	$\langle {\eta_1} ^2 \rangle$&	$\zeta_1$& $\zeta_{1th}$&Ratio\\
\hline
$16 \times 16$&	0.45&	0.20& 0.04 &	2.43&	0.97&0.397\\
$32 \times 32$&	0.45&	0.20& 0.04 &	2.43&	1.00&0.411\\
$36 \times 36$&	0.45&	0.20& 0.04 &	2.43&	1.01&0.416\\
$64 \times 64$&	0.45&	0.20& 0.04 &	2.43&	1.05&0.432\\
\hline
		\end{tabular}
	\end{center}
\end{table*}

\begin{table*}[h]
	\caption{Noise characteristics of the \cite{foucaut04} experiments in a turbulent flow: High magnification}
	\label{withflow2}       
	\begin{center}
		\begin{tabular}{lllllll}
			\hline			
			$N_p$ =&0.025&&&&&\\
			IW&	Pixels&	N&$W_1$&	$\sigma_1$&	$\sigma_{1th}$& Ratio\\
			\hline
			$32 \times 32$ &1024&	25.6&0.819&	0.184&	0.162&0.879\\
			$60 \times 60$&3600&	90  &0.903&	0.098&	0.091&0.924\\
			$64 \times 64$&4096&	102.4&0.909&	0.092&	0.085&0.926\\
			\hline
			IW&	$u'_1$&	${u'_1}^2$&	$\langle {\eta_1} ^2 \rangle$& $\zeta_1$&	$\zeta_{1th}$& Ratio\\
			\hline
			$32 \times 32$&	0.87&	0.76&	0.04& 	5.8&	4.49&0.774\\
			$60 \times 60$&	0.87&	0.76&	0.04& 	5.8&	4.95&0.854\\
			$64 \times 64$&	0.87&	0.76&	0.04& 	5.8&	4.98&0.859\\
			\hline
		\end{tabular}
	\end{center}
\end{table*}

\begin{table*}[h]
	\caption{Noise characteristics of the \cite{foucaut04} experiments in a turbulent flow: High dynamic range}
	\label{withflow3}       
	\begin{center}
		\begin{tabular}{lllllll}
			\hline
			$N_p$ =&0.025&&&&&\\
			IW&	Pixels&	N&$W_1$&	$\sigma_1$&	$\sigma_{1th}$&Ratio\\
			\hline
			$24 \times 24$ &576&	14.4&0.375&	0.359&	0.356&0.991\\
			$44 \times 44$&1936&	48.4&0.659&	0.196&	0.257&1.331\\
			$64 \times 64$&4096&	102.4&0.766&	0.135&	0.191&1.412\\
			\hline
			IW&	$u'_1$&	${u'_1}^2$&	$\langle {\eta_1} ^2 \rangle$& $\zeta_1$&	$\zeta_{1th}$&Ratio\\
			\hline
			$24 \times 24$&	2.2&	4.84&	0.00& 	12.40&	12.22&2.50\\
			$44 \times 44$&	2.2&	4.84&	0.00& 	12.40&	21.47&2.50\\
			$64 \times 64$&	2.2&	4.84&	0.00& 	12.40&	24.94&2.50\\
			\hline
		\end{tabular}
	\end{center}
\end{table*} 
Looking at the low magnification case, one can see that both $\sigma$ and $\zeta$ are underestimated (the ratio is about 0.65 for $\sigma$ and 0.40 for $\zeta$). This could be explained by the fact that the light sheet is very thin (0.5 mm that is about 4 wall units) and very close to the wall ($y^+ = 30$). The wall normal velocity fluctuations being maximum at that wall distance, there may be a significant out-of-plane loss of pairs. As $N_{p}$ was evaluated by looking at the single exposed images, it is probable that only part of the visible particle images contribute effectively to the correlation (dividing $N_{p}$ by 2 gives a perfect prediction). Nevertheless, the ratio is independent of the IW size meaning that the noise varies effectively as $1/N$. 

The situation is more favourable for the high magnification case. The turbulence intensity is not provided in the article for this case. As it is the same flow conditions as the low magnification one, we could deduce it based on the change of magnification and of PIV $\delta t$. One fact could explain this better result: the $\delta t$ has been reduced from $150$ to $100 \mu s$ between the two experiments with the same light sheet thickness. The out-of-plane loss of pairs is probably much less. Again, the ratio of $\sigma_1 / \sigma_{1th}$ is independant of the IW size.

Looking now at the high dynamic case, the agreement stays reasonable, especially for the 24x24 IW. As the dynamic range is 15 px in this case, the $W_1$ parameter has a significant role, especially for the smallest interrogation window as the overlap between the two windows is quite limited. This points out the fact that if the dynamic range is large enough and the interrogation window small enough, one could reach the stage where $W_1 = 0$. In the present case, an IW of 16x16 would do the job.

\subsection{Overall assessment of data}

Looking globally at tables \ref{noflow} to \ref{withflow3}, and putting aside the specificity of the Pulnix camera along $x_1$, one can conclude first that the noise variance varies effectively as $1/N$;  and second that based on a quantization error $\Delta = 0.5$, the proposed model provides a reasonable prediction of the noise level in 2D2C PIV measurements. This noise has essentially two origins: the velocity fluctuations inside the interrogation volume and the quantization error on the particle images. One striking result is that, thanks to the fact that $\Delta$ is squared and divided by $12$, the noise due to the velocity fluctuations inside the interrogation volume generally dominates the quantization error. Another important observation is that it is possible to minimize the noise in the direction of the mean flow by adjusting the Dynamic Range and the size of the interrogation window. This should have significant consequences in terms of the optimization of a PIV set-up used to measure turbulence, especially when looking at small scales.

In summary, the experimental results are quite supportive of the theory, especially given that they were obtained by assuming the hot-wire spectra to be the same as the volume-averaged  PIV one.  As noted earlier, this can not be exactly true due to the different spatial filters. These differences probably account partly for the noise ratios of theory to experiment being different from unity. In their paper, \cite{foucaut04} propose an approach based only on the PIV spectrum which would be worth investigating further. 

Another parameter which is probably not very accurate in the experiment is the particle concentration as given by $N_p$. At that time, these were estimated from the PIV images in a manner which is not very accurate (particle images were counted by eye in a few records to get an estimation of their concentration). In all cases, it would be easy to make the ratios very close to 1 just by adjusting $N_p$. The present theory puts emphasis on the importance of this parameter and calls for a more accurate determination of it. For example, the approach of \cite{warner14} could be useful.  An alternative possibility would be to use digital image processing directly on the PIV images following \cite{serra82}.

\section{Summary and conclusions}
In the present contribution, a generalization of the theory developed by previous authors (and presented for example by \cite{adrianwesterweel2011}) has been proposed. It explicitly calculates the PIV noise. The two original ideas used to develop this new theory were the use of generalized functions and of Lagrangian velocities.  Based on this approach it was possible to propose explicit formulas for the noise affecting the two-point correlation tensor, the single-point Reynolds stresses, the three-dimensional and one-dimensional spectra. The formulas obtained are in full agreement with those proposed by \cite{foucaut04} using an heuristic approach. 

The experiments provided by \cite{foucaut04}, which are used to validate the theory, were performed with very early PIV cameras. The Pulnix camera was still using line transfer (but not interline) and the Kodak camera was a CCD photographic camera providing double exposure images. Despite their limitations, thanks to these experiments it was possible to provide support to the present theory for both no-flow and with-flow cases. The overall agreement obtained is encouraging and provides some insight into the noise origin and behaviour in a PIV image. Both the velocity fluctuations inside the interrogation window and the quantization error on the particle images appear to have a significant effect, with the former usually dominating. The effect of both on the noise variance is only attenuated as $1/\sqrt{N}$  (which dies off relatively slowly), where N is the number of particle images inside the interrogation window. 

The primary value of the present results is surely that by bringing a better understanding of the noise origin they should permit a better optimization of the recording parameters of PIV experiments for the purpose of minimizing the noise on turbulence measurements. From this point of view, it is important to notice that the main contribution to this noise is from the velocity fluctuations inside the interrogation volume. To reduce this part, one can  increase the seeding concentration; but the effect, as mentionned above, goes as $\sqrt{N}$. The other solution is to reduce the interrogation volume size (but keeping constant the number of particles in it, which means also increasing the seeding concentration). In any case, an accurate estimation of $N$ using methods like the one suggested by \cite{warner14} would help predict and minimize the noise. In the direction of the mean flow, a careful adjustment of the Dynamic Range can also help to reduce the noise on the corresponding Reynolds stress.

An interesting question is raised in the appendices as whether using the displacement peak of the correlation alone to characterize the volume-averaged velocities is enough.  The results of this paper are only unbiased if weighted by the average number of particles over many realizations, and not by the instantaneous number of particles in each volume.   Further research is necessary to test whether this requires an unnecessarily large number of particles, when fewer might be used to advantage.

Due to the fact that PIV cameras have made considerable progress since the study by \cite{foucaut04}, it would be of high interest to repeat their experiments with an up-to-date camera and with a systematic variation and careful characterization of the particle concentration $N_p$. Such an experiment both in a fluid at rest and in a turbulent flow should allow a better understanding and quantification of the PIV noise. Such a study, combined with the use of simulations with an up-to-date synthetic image generator, could provide significant insight in the origin of the noise in PIV measurements of turbulent flows and ways to minimize it.

\section*{Acknowledgements}
The authors would like to acknowledge the contributions of Pr. J-M Foucaut of Ecole Centrale de Lille who participated in the original experiments of \cite{foucaut04} and was quite helpful to acquiring and understanding the data.    The theoretical portion of this work was mostly performed while WKG was Visiting Researcher at LML in 2009, funded by the ``R\'{e}gion Nord Pas de Calais".   It benefited  greatly from collaboration with Dr. Maja W\"{a}nstr\"{o}m during her Ph.D. studies at Chalmers University of Technology. CNRS and Centrale Lille are also acknowledged for  supporting financially the subsequent visits of Pr. George.\\

\section*{Appendix A:  The statistics of $1/n(\vec{y},t)$ \label{sec_app_n}}
The conventional wisdom in PIV is that the average number of
particles in the interrogation volume must be chosen to at least 5
to provide reliable signals (Raffel et al.\ \cite{raffel98,raffel2011}), where more is
generally considered to be better (for reasons which will be obvious
below).
Since $N$ is assumed to be at least marginally much greater than unity, we can use this to
our advantage to develop an expansion of $1/{n}(\vec{y},t)$ in
powers of $n'/N$ where $n'(\vec{y},t)$ is defined to be the instantaneous
fluctuation in the number of particles in the volume; i.e.,

\begin{eqnarray}
n'(\vec{y},t) & = & {n}(\vec{y},t) - N
\end{eqnarray}
It follows that:
\begin{eqnarray}
\frac{1}{{n}(\vec{y},t)} & = & \frac{1}{ N\left[{ 1 +
		{n'(\vec{y},t)}/{N} }\right]}
\end{eqnarray}

Binomially expanding the denominator in powers of $n/N$ yields:
\begin{equation}
\frac{1}{n(\vec{y},t)} = \frac{1}{N} \left[{1-
	\frac{n'(\vec{y},t)}{N} + \left(\frac{n'(\vec{y},t)}{N}\right)^2 -
	\cdots }\right] \label{eq:inversenexpansion}
\end{equation}
It follows immediately that, at first order:

\begin{equation}
\langle {{\frac{1}{n(\vec{y},t)}}}\rangle =\frac{1}{N}
\left[ 1 + \frac{\langle n'^2(\vec{y},t) \rangle}{N^2} + \cdots
\right]
\end{equation}
where $\langle n^2(\vec{y},t) \rangle = var\{\tilde{n}(t)\}$.

From the definitions of $n$ and $N$:

\begin{equation}
n(\vec{y},t) = \int_{a s} \left[ g(\vec{a}) -  \mu \right]  w(\vec{y}, \vec{X}[\vec{a},t]-\vec{y}) d\vec{a}
\end{equation}
It follows that the variance of $n$ is given by:

\begin{equation}
var\{n(\vec{y},t)\} = \int\int_{a s} \langle  
w(\vec{y},\vec{X}-\vec{y}) 
w(\vec{y},\vec{X'}-\vec{y})g(\vec{a})g(\vec{a'})\rangle
d\vec{a} d\vec{a'}  - N^2
\end{equation}

From before (\cite{GeorgeLumley1973,Buchhaveetal1979,velte09}) we already know that the two-point correlation of the $g$'s is
given by:

\begin{eqnarray}
\langle g(\vec{a})g(\vec{a'}) \rangle = \mu^2 +
\mu \delta(\vec{a'}
- \vec{a}) \label{eq:ggprime}
\end{eqnarray}
This follows from the fact that any two $g(\vec{a})$'s
are statistically independent and therefore uncorrelated, so:

\begin{equation}
\langle [g(\vec{a}) - \langle g(\vec{a})][g(\vec{a'})- \langle
g(\vec{a'}) ]\rangle = \mu \delta(\vec{a'} - \vec{a}).
\end{equation}

Using equation~(\ref{eq:ggprime}) and transforming as above from
Lagrangian to Eulerian coordinates yields:

\begin{eqnarray}
var\{n\}  =  & & \mu^2 \left[\int_{a s} \langle w(\vec{y},\vec{X}
- \vec{y})\rangle d\vec{a} \right]^2 
+ \mu \int_{a s} \langle w(\vec{y},\vec{X}
- \vec{y})
\rangle^2 d\vec{a}  - N^2 \nonumber \\
= & & \mu^2 \left[\int_{a s} w(\vec{y},\vec{X}
- \vec{y}) d\vec{x} \right]^2  + \mu \int_{a s}  [w(\vec{y},\vec{X}
- \vec{y})]^2 d\vec{x} - N^2
\end{eqnarray}
But the first integral is just $(\mu V)^2 = N^2$.  Similarly the
second integral is just $N$, since $w^2 = w$ given its description
as off-on in equation~(\ref{eq:volume}).  Thus the variance of the
number of particles is equal to the mean number of particles.

\begin{equation}
var\{{n}(\vec{y},t)\} = N
\end{equation}
Note this is quite independent of any assumptions about the
probability density function of the particle distribution.
\footnote{It is commonly assumed that the particles are Poisson
	distributed, without much concern for whether their statistics are
	correlated with the flow.  Here we have derived that they are
	indeed Poisson-distributed to at least second order (and probably
	all orders) since the mean and variance are equal.  And we have done
	so without any assumptions about how they relate statistically to
	the flow, a consequence of our assumptions about how the particles
	were distributed initially.}

So in summary, to first order in $1/N$,
the expected value of $\langle 1/n(\vec{y},t) \rangle $ is:

\begin{equation}
\langle \frac{1}{n(\vec{y},t)} \rangle = \frac{1}{N} ~\left[1 + \frac{1}{N} \right]
\label{eq:navg-expansion}
\end{equation}

\section*{Appendix B:  The biased `output' velocity, ${u}_{ooi}(\vec{y},t)$ \label{app-B}}

In this appendix we will use the nomenclature, ${u}_{ooi}(\vec{y},t)$ for the biased velocity to distinguish it from the unbiased ${u}_{oi}(\vec{y},t)$.  The basic idea is that we should be dividing by the ${n}(\vec{y},t)$ which is the actual number of particles in the volume at an instant.  Interestingly, this turns out to introduce a bias, in part for reasons clear from Appendix A,

The intuitive idea is that the `output velocity', say
${u}_{ooi}(\vec{y},t)$ from a particular interrogation window
centered at location, $\vec{y}$, should be represented using
generalized functions as an integral over {\it particle} initial
coordinate in space; i.e.,

\begin{equation}
{u}_{ooi}(\vec{y},t) = \frac{1}{{n}(\vec{y},t)}\int_{as} {v_i}(\vec{a},t) w(\vec{y},\vec{X}[\vec{a},t]-\vec{y})
g(\vec{a}) d\vec{a} \label{eq:basic1}
\end{equation}

Except for the factor of
$1/{n}(\vec{y},t)$ in front of the integral (instead of $1/N$), this is identical to equation~\ref{eq:basicunbiased}.  So it is in effect the `{\it arithmetic average}' over the particles in the volume at an instant in time.  By contrast, equation~\ref{eq:basicunbiased}, is an integral sum of the contributions of the particles present multiplied by $1/N$, the average value of $N$ over all realizations.

Averaging equation~\ref{eq:basic1} yields:

\begin{eqnarray}
\langle {u}_{ooi}(\vec{y},t) \rangle &  = & \langle { \frac{1}{{n}(\vec{y},t)}\int_{as} {v_i}(\vec{a},t) w(\vec{y},\vec{X}[\vec{a},t]-\vec{y})
	g(\vec{a}) d\vec{a} } \rangle  \label{eq:basicavg} \\
& = & \langle  \frac{1}{{n}(\vec{y},t)} \rangle ~\langle {\int_{as} {v_i}(\vec{a},t) w(\vec{y},\vec{X}[\vec{a},t]-\vec{y})
	g(\vec{a}) d\vec{a} }\rangle  \nonumber
\end{eqnarray}
where we have used the statistical independence of the particle locations and their motion in initial position coordinates to split the integrals.\footnote{Note this can also be shown directly by using the defining integral for $n(\vec{y},t)$ together with the Taylor expansion for $1/n(\vec{y}t)$.}

But this integral is just our volume-averaged velocity (multiplied by $N$). Using equation~\ref{eq:navg-expansion} implies immediately that to first order in $1/N$, the biased velocity is:

\begin{equation}
\langle {u}_{ooi}(\vec{y},t) = \langle
\tilde{u}_i(\vec{y},t) \rangle \left[{1 + \frac{1}{N} }\right]
\end{equation}
{\bf Clearly this is a biased mean value toward higher values. And it 
	is not particularly small, even for values of $N$ near 10.}  
The same arguments can be applied to the second moments with the consequence that all need an additional factor of $1+3/N$.

\bibliographystyle{unsrt}

\bibliography{biblio_pivwkg}

\end{document}